\documentclass[a4paper]{quantumarticle}
\pdfoutput=1
\usepackage{graphicx}
\usepackage{amsmath}
\usepackage{amssymb}
\usepackage{mathtools}
\usepackage{braket}
\usepackage{booktabs}
\usepackage{multirow}
\usepackage{hyperref}
\usepackage[numbers,sort&compress,square]{natbib}
\usepackage{subcaption}

\newcommand{\tablelcell}[2]{\parbox[c]{#1}{\setlength{\baselineskip}{1.12\baselineskip}\raggedright\hyphenpenalty=10000\exhyphenpenalty=10000 #2}}
\newcommand{\tableccell}[2]{\parbox[c]{#1}{\setlength{\baselineskip}{1.12\baselineskip}\centering\hyphenpenalty=10000\exhyphenpenalty=10000 #2}}

\newcommand{\sSz}{{}^1\mathrm{S}_0}
\newcommand{\tSo}{{}^3\mathrm{S}_1}
\newcommand{\sPo}{{}^1\mathrm{P}_1}
\newcommand{\tPz}{{}^3\mathrm{P}_0}
\newcommand{\tPo}{{}^3\mathrm{P}_1}
\newcommand{\tPt}{{}^3\mathrm{P}_2}
\newcommand{\tDo}{{}^3\mathrm{D}_1}
\newcommand{\fermi}{{}^{171}\mathrm{Yb}}
\newcommand{\boson}{{}^{174}\mathrm{Yb}}

\begin{document}

\title{Quantifying the Dual-isotope Advantage for Ytterbium-array Surface Codes using Realistic Noise Models}

\author{Fumiyoshi Kobayashi}
\affiliation{R4D, Mercari, Inc., 6-10-1 Roppongi, Minato-ku, Tokyo, 106-6126, Japan}

\author{Toshi Kusano}
\affiliation{Department of Physics, Graduate School of Science, Kyoto University, Kyoto 606-8502, Japan}

\author{Nicholas Fazio}
\affiliation{Yaqumo Inc., 3-2-2 Marunouchi, Chiyoda-ku, Tokyo 100-0005, Japan}

\author{Yuma Nakamura}
\affiliation{Yaqumo Inc., 3-2-2 Marunouchi, Chiyoda-ku, Tokyo 100-0005, Japan}

\date{August 28, 2026}

\begin{abstract}
  Neutral-atom quantum computers are a promising platform for fault-tolerant quantum computation, but logical performance depends on systemic realistic noise factors during syndrome extraction. In dual-isotope Yb arrays, the roles of data and ancilla qubits are separated spectrally, allowing ancilla qubits to be measured in place without additional transport or shelving operations. Here we quantify the advantage of a dual-isotope Yb architecture for surface code memories. We develop an experimentally motivated Clifford-compatible noise model for dual-isotope $^{171}$Yb--$^{174}$Yb systems using generalised Pauli twirling and implement it as a wrapper for \textit{Stim} called \textit{DualYbSim}~\cite{DualYbSim}, which has been packaged as an open source Python library. Simulations of rotated and XZZX surface codes show that a dual-isotope architecture with in-place measurement achieves the lowest logical error rates among the architectures considered, outperforming single-isotope schemes based on shelving or zoned measurement. Our error-budget analysis also identifies Rydberg-state decay as the dominant limitation, contributing to 74-80~\% of the logical error rate scaling, highlighting concrete experimental targets for improving FTQC performance.  
\end{abstract}

\maketitle

\section{Introduction}
\label{sec:introduction}
\quad Neutral atoms in optical tweezer arrays have emerged as a promising platform for quantum computing~\cite{saffman2010,saffman2016,henriet2020,saffman2025}, offering advantages such as long coherence times~\cite{young2020half,Jenkins2022,barnes2022,Yang2025Minute}, high-fidelity gate operations~\cite{madjarov2020,Evered2023,Peper2024Spec,Tsai2024Bench,Infleqtion2024Univ,AC2025High,Senoo2025High,lib2026velocity,evered2026high,liu2026high}, scalability~\cite{ebadi2021,scholl2021,Schymik2022situ,Lars2024super,Tao2024lattice,Gyger2024Cont,Norcia2024Ite,Manetsch20246100,Pichard2024Cryo,Li2025Fast,Chiu2025Continuous,zhu2025high,Holman2026,wang2026_11000}, and flexible connectivity enabled by coherent atom transport~\cite{beugnon2007,bluvstein2022,Manetsch20246100}. These features constitute a versatile framework for exploring resource-efficient quantum error correction (QEC) schemes and logical operations, including high-rate QEC codes~\cite{Bravyi2024BB,Goto2024,Yoshida2025C4C6,Xu2024Constant,tamiya2026fault,Pecorari2025LDPC,Poole2025LDPC,zhao2026ultra} and transversal logic~\cite{bluvstein2024logical,Chain2024,Sahay2025,zhou2025low,cain2025fast,Marc2026,Turner2026}.

While the potential of neutral-atom processors for fault-tolerant quantum computing (FTQC) is widely recognised, evaluating bottlenecks in their QEC performance is crucial for implementing feasible FTQC. When compared to superconducting qubit systems, neutral-atom processors are more constrained by slower operational speeds. 
In particular, although the measurement overhead of neutral-atom processors can be mitigated through fast imaging~\cite{su2025fast,Falconi2025,yokoyama2026} and algorithmic fault tolerance~\cite{zhou2025low}, a significant contribution to the time overhead remains as atom transportation, limiting the execution time of general logical circuits~\cite{Zhou2025Resource,sunami2025}.

Dual-species (or dual-isotope) systems~\cite{Sheng2022,Singh2022,yuma2024,weber2026dual} can mitigate this runtime overhead by in-place measurement of ancilla qubits, avoiding the transportation of ancilla qubits to a dedicated measurement zone. Due to the spectral separation between isotopes, readout crosstalk to data qubits can be suppressed even when the readout laser is applied to the entire array. This enables constant-time syndrome extraction in each QEC cycle, independent of code distance~\cite{sunami2025}, such that logical time overheads remain scalable. Recent experimental progress, including defect-free array generation~\cite{Sheng2022,Singh2022,yuma2024,wei2025}, low-crosstalk readout~\cite{singh2023mid,yuma2024}, and inter-species two-qubit gates~\cite{Zeng2017,anand2024dual,white2026,miles2026,wang2026stab}, establishes dual-species atom arrays as a compelling candidate for scalable neutral-atom quantum computing. However, quantitative studies of dual-species systems for QEC remain limited, particularly those comparing their performance with single-species configurations that employ atom transportation or shelving during syndrome extraction. 

In this work, we methodically consider realistic sources of noise on a neutral-atom device to evaluate the performance of a quantum memory on the rotated and XZZX surface codes~\cite{RotatedSurfaceCode2007-Bombin, XZZX2021} for several distinct realisations utilising dual-isotope ytterbium (Yb). Using the noise model developed for this work, which has been packaged as a publicly available open-source Python library~\cite{DualYbSim}, we simulate and compare the logical error rates across various Yb-isotope configurations including single-isotope configurations with atom transportation~\cite{Bluvstein2025-ge,Muniz2025Repeated,zhang2025lev,AC2026toric} or shelving~\cite{Lis2023} during syndrome extraction (SE).
Our results demonstrate that a dual-isotope approach utilising $\fermi$ ground-state qubits and $\boson$ optical clock-state qubits as data and ancilla qubits respectively achieves the smallest logical error rate (LER) and highest error suppression factor compared to other configurations, such as single-isotope configurations that utilise atom transportation or shelving during syndrome extraction.
Furthermore, we numerically analyse the contribution of physical errors to the logical error suppression factor, showing that approximately 74 to 80~\% of the error contribution comes from Rydberg decay. We also investigate the impact under reduced Rydberg decay, which suggests that the error suppression factor would be enhanced by a factor of 1.7 if the decay rate from the Rydberg state could be halved. Moreover, this enhancement would be even greater if the depolarising errors are simultaneously reduced, even slightly.
These results demonstrate some of the advantages of dual-Yb systems compared to single-isotope systems, and identify them as sensible candidates for performing QEC with surface codes on current and upcoming hardware.

This paper is organised as follows. Section~\ref{sec:noise_model} describes the dual-Yb noise model, including qubit encoding and noise channels. Section~\ref{sec:surface_code} details the surface code construction and numerical simulation setup. In Sec.~\ref{sec:results}, we present the simulation results and an error budget analysis. Finally, Section~\ref{sec:conclusion} summarises our findings and discusses future directions.

\section{Noise Model of the Dual Ytterbium System}
\label{sec:noise_model}
\quad This section presents the noise model for the dual-isotope Yb platform used in our simulations. We begin by describing the qubit encodings and Rydberg-gate pathways for $\fermi$ and $\boson$, introducing effective multi-level models that include effective loss states. We then summarise the assumed error channels for each operation type. Finally, to make large-scale QEC simulation tractable, we map the resulting non-Pauli physical noise to a Clifford-simulatable form using generalised Pauli twirling~\cite{DualYbSim, Google-Quantum-AI2023-yd, gidney2021stim}.

\subsection{Dual-isotope Ytterbium qubits}
\label{subsec:dual_yb_qubit}

\begin{figure}[t]
  \centering
  \includegraphics[width=\columnwidth]{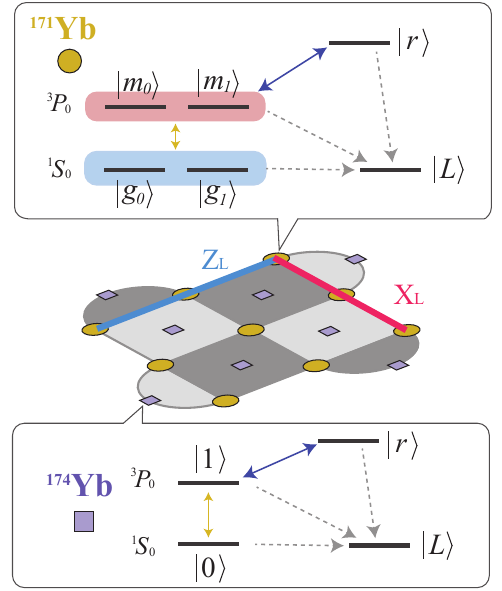}
  \caption{\textbf{Schematic of the surface code layout in a dual-isotope Yb system.} Data qubits (yellow circle) and ancilla qubits (purple square) are encoded in $\fermi$ and $\boson$, respectively. For the $\fermi$ data qubits, we evaluate two encoding schemes: the $\sSz$ ground-state encoding (g-qubit: $\ket{g_\bullet}$, shaded in blue) and the $\tPz$ metastable-state encoding (m-qubit: $\ket{m_\bullet}$, shaded in red). The $\boson$ ancilla qubits are encoded as optical clock qubits, with $|0\rangle$ and $|1\rangle$ representing the ground and metastable states. Rydberg gates are mediated by the Rydberg blockade effect via a high-lying Rydberg state $|r\rangle$. All computational states are susceptible to loss, depicted as grey dashed arrows; the state $|L\rangle$ encompasses various loss channels, including out-of-trap loss and decay into dark states such as the $\tPt$ state.}
  \label{fig:surface_code_geometry}
\end{figure}

\quad Ytterbium (Yb) is an alkaline-earth-like atom with a rich internal structure, making it a versatile candidate for quantum computing~\cite{Saskin2019,Jenkins2022,Ma2022,Okuno2022,Norcia2023,Huie2023,Falconi2025,zhu2025high}. The two valence electrons in Yb give rise to a variety of electronic states, including the zero-total-angular-momentum ground state ($\sSz$) and long-lived metastable states (e.g.~$\tPz$). These states can be used to encode qubits with distinct advantages, such as long coherence times~\cite{Jenkins2022,Lis2023}, the capability for mid-circuit operations~\cite{Lis2023,Norcia2023,Huie2023,Ma2023high}, and high-fidelity Rydberg excitation~\cite{Peper2024Spec,Senoo2025High,AC2025High,liu2026high}. Furthermore, the existence of multiple isotopes with distinct nuclear spins~\cite{Jenkins2022,Ma2022,Okuno2022,zhu2025high,Abdel2025,kusano2026} allows for the implementation of dual-isotope systems, where different isotopes serve as data qubits and ancilla qubits~\cite{yuma2024}.
Here, we focus on a quantum processor composed of $\fermi$ and $\boson$ atoms, where the former is employed as data qubits and the latter as ancilla qubits (see Fig.~\ref{fig:surface_code_geometry} and related energy-level diagrams depicted in Fig.~\ref{fig:energy-level_diagram}).

\paragraph{Data-qubit encoding.} There are two primary schemes for encoding the $\fermi$ data qubit. The first utilises the nuclear-spin degrees of freedom in the $F=1/2$ manifold of the $\sSz$ ground state~\cite{Jenkins2022,Ma2022}, denoted as a $\fermi$-g qubit spanned by $|g_0\rangle$ and $|g_1\rangle$. The second utilises the $F=1/2$ manifold of the metastable $\tPz$ state~\cite{Lis2023,Ma2023high}, denoted as a $\fermi$-m qubit spanned by $|m_0\rangle$ and $|m_1\rangle$. In the g-qubit scheme, performing two-qubit gates requires an excitation to the $\tPz$ manifold~\cite{AC2025High}; thus, the two-qubit gate error inherently includes $\tPz$ excitation error. In contrast, the m-qubit scheme uses the $\tPz$ manifold as the computational subspace, eliminating the need for this excitation and potentially offering higher gate fidelity. However, unlike the g-qubit, the m-qubit is susceptible to errors from decay to the ground state induced by nuclear-spin control lasers or trapping-tweezer beams~\cite{Senoo2025High, zhang2025lev, Siegel2024}. Thus, it is not immediately clear which encoding yields superior noise performance. In this study, we compare the LERs of the surface code for both g-qubit and m-qubit encodings within a dual-isotope system.

\paragraph{Ancilla-qubit encoding.} For $\boson$, we utilise an optical clock qubit as the ancilla qubit, encoded in the ground state $\sSz$ ($|0\rangle$) and the metastable $\tPz$ state ($|1\rangle$). Unlike the nuclear-spin qubits of $\fermi$, which operate within a nearly energy-degenerate manifold, the energy splitting of the $\boson$ clock qubit exceeds several hundred THz. This large separation allows non-destructive, state-selective readout using lasers resonant with the $\sPo$ or $\tPo$ transitions: fluorescence is obtained only when the atom is in the addressed manifold, enabling bright/dark discrimination between qubit states. To implement loss-aware readout~\cite{Norcia2023,Senoo2025High,Baranes2026,liu2026Loss}, we employ a three-step procedure: (1) fluorescence detection via the $\sPo$ or $\tPo$ transition to identify atoms in the ground state, (2) a $\pi$-pulse on the $\sSz \leftrightarrow \tPz$ clock-transition to swap the populations of $|0\rangle$ and $|1\rangle$, and (3) a second fluorescence detection to identify atoms originally in the metastable state. This sequence allows us to distinguish between $|0\rangle$, $|1\rangle$, and loss events.
However, note that we don't make use of atom-loss flags in this work, simplifying our decoding considerations.

\paragraph{Two-qubit gate pathway.} Two-qubit gates are implemented via the Rydberg blockade effect~\cite{saffman2010}, where the excitation of one atom to a Rydberg state prevents the excitation of a neighbouring atom, thereby enabling conditional gate operations. For the $\fermi$-g qubit, the $\tPz$ excitation is required to access the Rydberg state, while for the $\fermi$-m and $\boson$ qubits, direct single-photon excitation is available. We denote the Rydberg state used for these operations as $|r\rangle$, typically chosen from a high-lying manifold such as $\tSo$~\cite{Peper2024Spec,Senoo2025High}.

We construct a dedicated noise model for each qubit type. Although neutral atoms possess complex sublevel structures, such as Zeeman sublevels, we adopt simplified schemes: a four-level system $\{|0\rangle, |1\rangle, |r\rangle, |L\rangle\}$ for the $\boson$ qubits, and a six-level system $\{|m_0\rangle, |m_1\rangle, |g_0\rangle, |g_1\rangle, |r\rangle, |L\rangle\}$ for the $\fermi$ qubits. Here, $|m_\bullet\rangle$ and $|g_\bullet\rangle$ denote the computational basis states of the metastable state and ground state qubits, respectively. The state $|L\rangle$ represents effective loss channels, including out-of-trap loss and decay into dark states such as the $\tPt$ state.

The noise channels are categorised into coherent-control, measurement, reset, idling, decay, and transportation channels, following the definitions in Appendix~\ref{app:noise_channel_definitions}. Here, the coherent-control category includes errors arising from single-qubit gates, two-qubit gates, and clock-transition excitation ($\sSz\leftrightarrow\tPz$). Clock-transition excitation is particularly notable as a source of noise, since it factors into the measurement of the $\tPz$ state, into Rydberg excitation for $\fermi$-g qubits, and into the shelving of $\fermi$ qubits. As described in Sec.~\ref{subsec:approximation_to_pauli_noise_model}, non-Pauli error channels such as decay are converted to a generalised Pauli representation for efficient Clifford simulation.

\subsection{Measurement architectures on neutral-atom quantum computers}

\quad Due to the diverse sublevels provided by the atoms and the versatility enabled by dual-isotope (species) systems of neutral-atom quantum computers, there are several possible methods for mid-circuit measurement. 

\paragraph{In-place measurement.} 
This kind of method is favourable for implementing mid-circuit measurement since it doesn't depend on atom transportation. In neutral-atom processors, the imaging beam used for qubit-state readout is typically applied globally to the entire array or to a selected region. When atoms are measured in place within the illuminated region, single-isotope systems generally require additional local operations to prevent untargeted qubits from fluorescing. In contrast, dual-isotope systems naturally enable in-place mid-circuit measurement without these additional local operations, because untargeted qubits of the other isotope are off-resonance with the measurement light~\cite{singh2023mid,yuma2024}. Thus, we adopt a dual-isotope system to implement `in-place measurement'.

\paragraph{Shelving.}
Another method is to shelve the qubit state before measurement.
To avoid resonance with the measurement laser, we can \textit{shelve} qubit sublevels to other off-resonance sublevels on the unmeasured qubits by exciting these sublevels site-selectively before the measurement~\cite{Lis2023}. 
This allows single-isotope systems to perform in-place measurements, however, it introduces additional operations before and after the measurement.
This method, for example, can be applied to a $\fermi$-g qubit because the $\fermi$-m qubit works as a \textit{shelf} of the $\fermi$-g qubit via the clock-transition, which isolates the shelved qubit from the fluorescence of $\fermi$-g qubits.
Although shelving is also achieved `in-place', we call it shelving throughout the text to distinguish it from the dual-isotope method. 

\paragraph{Zoned measurement.}
The final technique utilises atom transportation to transfer qubits to measurement zones, such as in zoned architectures~\cite{bluvstein2022,bluvstein2024logical,AC2026toric}.
While the qubits in single-isotope systems can be measured in a separated zone, this mandates the transportation of qubits to a distant zone and associated handover operations from the spatial light modulator, which holds atoms in place, to the acousto-optic deflector, which moves them to the measurement zone and the reset zone.

\vspace{10pt}
After measurement the ancilla qubits are reset to $\ket{0}$ and depending on the measurement outcome they are replenished by a fresh atom from the atom reservoirs, necessitating atom transportation.
In this work, we assume the qubits measured in the measurement zone are reset in that zone before returning back to the storage zone, while the qubits measured via in-place measurement are replenished and reset directly in the storage zone. We ignore transportation noise associated with replenishment from the atom reservoirs, which is common to all configurations, since we do not make use of atom-loss flags or other complexities related to replenishment.

\subsection{Approximating Noise Channels as Pauli Noise}
\label{subsec:approximation_to_pauli_noise_model}

\quad Atom qubits encounter many sources of noise, described in Appendix~\ref{app:noise_channel_definitions}.
These noise channels include noise that is fundamentally non-Pauli in nature and thus cannot be efficiently simulated on a classical computer because the exact simulation requires heavy simulation of density matrices.
Pursuing exact state evolution is computationally expensive, however, by utilising the Pauli twirling approximation~\cite{Geller2013-tn}, we can convert noise represented by a set of Kraus operators into Pauli noise channels, which enables classical tableau simulation of stabiliser states~\cite{Aaronson2004-oi}.

In the case of neutral-atom qubits, they consist not only of qubit states $\{\ket{0}, \ket{1}\}$ or $\{\ket{\bullet_0}, \ket{\bullet_1}\}$, but also of other states that serve as a source of leakage or loss such as  $\ket{r}$ or $\ket{L}$.
To incorporate population transfer to these sublevels as well, we employ the generalised Pauli twirling approximation described in the supplementary information of Ref.~\cite{Google-Quantum-AI2023-yd}.
This converts the Kraus operators performed on the larger Hilbert space, including the effective loss states, into two divorced error processes, i.e. Pauli error channels and erasure channels with their own respective probability distributions. Note well, the erasure channels produced by this approximation are not always heralded. Since we don't use such information for decoding, in this work we refer to these channels collectively as erasure, which includes sources of both leakage and loss.

To perform generalised Pauli twirling on the noise channel of each gate operation, we first determine the Kraus operators $\{K_{j>0}\}$ of those noise channels, representing each channel as $\mathcal{E}_N(\rho) = \sum_{j} K_j \rho K^{\dagger}_j$.
Using the generalised Pauli twirling approximation on $\mathcal{E}_N$, we can obtain the erasure and Pauli error probabilities for each gate operation: $\{(\mathbb{P}(i\to f),\; \mathbb{P}(P_{\mu}|i \to f))\}_{i,f}$.
Here, $\mathbb{P}(i\to f)$ is the transition probability from the initial subspace~$i$ to the final subspace~$f$ through the operation of some $K_j$, and $\mathbb{P}(P_{\mu}|i \to f)$ is the corresponding conditional probability of the Pauli operator $P_{\mu},\;\mu\in\{0,1,2,3\}$ acting on the qubit via the transition.

Now consider the qubit subspace $\mathcal{H}_{c}=\{\ket{0}, \ket{1}\}$ and the subspace $\mathcal{H}_{e}$ orthogonal to it, and projectors $\Pi_c = \ket{0}\bra{0} + \ket{1}\bra{1}$ into the qubit space and $\Pi_e$ into the rest.
While erasure errors occur during the transition $c\to e$, the error on the qubit subspace is induced by the transition $c\to c$.
Then the Kraus operator acting on $\mathcal{H}_{c}$ is given by $K^{c}_j = \Pi_{c} K_j \Pi_{c}$.
$K^{c}_j$ can be expanded as a linear combination of Pauli operators:
\begin{equation}
    K^{c}_j = \sum_\mu a^{j,c}_\mu P_{\mu}.
\end{equation}
The probability of applying $P_\mu$ in a twirled Pauli channel of~$\mathcal{E}_N$ is given as
\begin{equation}
    p_\mu = \sum_{j}|a^{j,c}_\mu|^2.
\end{equation}
While the Pauli error probabilities are given above, the transition probability is given by
\begin{equation}
    \mathbb{P}(c\to e) = \sum_{j, c, e}|\bra{e}K_j\ket{c}|^2,
\end{equation}
where $\ket{c}$ and $\ket{e}$ are the orthogonal bases of $\mathcal{H}_c$ and $\mathcal{H}_{e}$, respectively.

Finally, the twirled Pauli error channel of each gate operation is directly implemented as 
\begin{align}
  \mathcal{E}_{\text{error}}(\rho) &\approx (\mathcal{E}_{\text{erase}}\circ \tilde{\mathcal{E}}_{\text{twirl}})(\rho),
  \label{eq:error_channel_approximation}
\end{align}
where the channel $\tilde{\mathcal{E}}_{\text{twirl}}$ is a twirled Pauli error channel with probability distribution $\{p_\mu\}$ for each term approximating to a Pauli error transition in the qubit subspace,
\begin{align}
  \tilde{\mathcal{E}}_{\text{twirl}}(\rho) &= \sum_{\mu=0}^3 p_{\mu} P_{\mu} \:\rho \:P_{\mu}^\dag, \quad \sum_{\mu=0}^3 p_{\mu} = 1,
  \label{eq:twirled_pauli_error_channel}
\end{align}
where $P_\mu$ are the Pauli operators, and the channel~$\mathcal{E}_{\text{erase}}$ is an erasure channel representing transitions to $e$ that never return to $c$ without the reset operation. Thus, it is treated as a completely depolarising channel with error probability equal to the decay probability~$p$ from the qubit subspace to the non-computational subspace.
\begin{equation}
 \begin{aligned}
  \mathcal{E}_{\text{erase}}(\rho, p)
  &= (1-p)\:\rho + p\:\frac{I}{2} \\
  &= (1-\frac{3p}{4})\:\rho
    + \frac{p}{4}\:\sum_{i=1}^3 P_i \:\rho \:P_i^\dag
 \end{aligned}
  \label{eq:erasure_channel}
\end{equation}
Each approximated error channel is detailed in Appendix~\ref{app:error_channel_approximation}.

\section{Surface Code Construction and Numerical Simulation Setup}
\label{sec:surface_code}

\quad The surface code is a quantum error-correcting code anticipated to be a promising candidate for near-term fault-tolerant quantum computing~\cite{Dennis2001-jj, Horsman2011-yv}.
Consider a $d \times d$ square lattice where data qubits are placed on the vertices, with ancilla qubits at the centre of the faces.
To define the stabiliser group for the surface code, we partition the faces into two sets using a checkerboard pattern, with elements $f_X \in F_X$ and $f_Z \in F_Z$.
The stabiliser operators are given by
\begin{equation}
  X_{f_X} = \bigotimes_{v \in \partial {f_X}} X_v, \quad
  Z_{f_Z} = \bigotimes_{v \in \partial {f_Z}} Z_v,
  \label{eq:stabilizers}
\end{equation}
where $f_X$ and $f_Z$ respectively denote the X- and Z- faces of the square lattice, $v$ is a vertex, and $\partial$ is the boundary operator returning the set of vertices for each face, although this slightly abuses the usual usage of boundary operators from algebraic topology.

\begin{figure}[htbp]
  \centering
  \begin{subfigure}{\columnwidth}
  \centering
  \includegraphics[width=0.8\columnwidth]{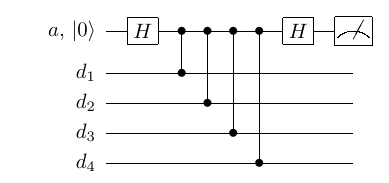}
  \caption{Z stabiliser check}
  \end{subfigure}
  \begin{subfigure}{\columnwidth}
  \centering
  \includegraphics[width=0.8\columnwidth]{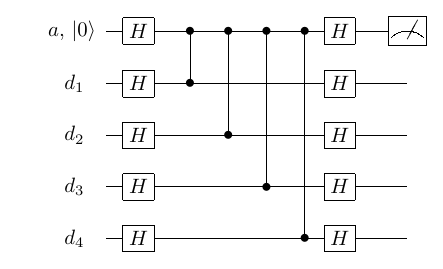}
  \caption{X stabiliser check}
  \end{subfigure}
  \newline
  \begin{subfigure}{0.28\textwidth}
  \centering
  \includegraphics[width=\textwidth]{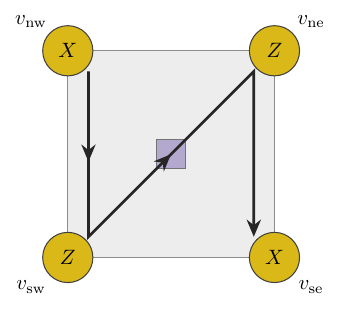}
  \caption{XZZX stabiliser check}
  \label{fig:xzzx_face}
  \end{subfigure}
  \caption{Standard syndrome extraction circuits for the surface code. CZ gates are a native two-qubit gate in neutral-atom quantum devices utilising Rydberg blockade. Panels (a) and (b) show the $Z$- and $X$-stabiliser check circuits of the rotated surface code, respectively. Panel (c) shows a single face of the XZZX surface code, where each data qubit (yellow circle) on the vertices $v_{\mathrm{nw}}$, $v_{\mathrm{ne}}$, $v_{\mathrm{sw}}$ and $v_{\mathrm{se}}$ is acted on by the Pauli operator denoted inside the circle and the N-shaped arrow denotes the order in which the CZ gates are applied between each data qubit and the ancilla qubit at the centre of a face. Since every face of the XZZX code carries a generator of the same form, the same ordering is used for all faces.}
  \label{fig:se_circuit}
\end{figure}

The logical qubit state is encoded into the $+1$ eigensubspace of these stabiliser operators.
The logical operators $X_L$ and $Z_L$ are defined as shown in Fig.~\ref{fig:surface_code_geometry}: $X_L$ is a chain of Pauli-$X$ operators traversing the lattice vertically across data qubits, and $Z_L$ is a chain of Pauli-$Z$ operators traversing horizontally.
These operators commute with all stabilisers in Eq.~\eqref{eq:stabilizers} and serve as the logical Pauli-$X$ and Pauli-$Z$ observables, respectively.
Errors on this code are detected as operators that anticommute with one or more stabilisers.
To detect these errors, we perform syndrome extraction (SE), which consists of indirect measurements of the stabiliser operators via circuits like shown in Fig.~\ref{fig:se_circuit}.
Decoding the detected error syndromes enables quantum error correction.

In addition to the rotated surface code above, we employ the XZZX surface code~\cite{XZZX2021}, a non-CSS variant defined on exactly the same lattice, with the same number of data and ancilla qubits and the same code distance $d$.
In contrast to Eq.~\eqref{eq:stabilizers}, the faces are no longer partitioned into $F_X$ and $F_Z$. Every face represents a single stabiliser generator whose Pauli type alternates around the face.
Labelling the vertices of a face by their position relative to its centre each XZZX generator is 
\begin{equation}
  S_f = X_{v_{\mathrm{nw}}} \otimes Z_{v_{\mathrm{ne}}}
        \otimes Z_{v_{\mathrm{sw}}} \otimes X_{v_{\mathrm{se}}},
  \label{eq:xzzx_stabilizers}
\end{equation}
where $v_{\mathrm{nw}}$, $v_{\mathrm{ne}}$, $v_{\mathrm{sw}}$ and $v_{\mathrm{se}}$ denote the positions of vertices on an XZZX face, as shown in Fig.~\ref{fig:se_circuit}(c), with $\partial f = \{v_{\mathrm{nw}}, v_{\mathrm{ne}}, v_{\mathrm{sw}}, v_{\mathrm{se}}\}$.
The XZZX code is local-Clifford equivalent to the rotated surface code by applying Hadamard gates to the data qubits placed diagonally on one sublattice of the checkerboard coloured faces $X_{f_X}$ and $Z_{f_Z}$.
The logical operators follow from the same conjugation, and are hence supported on the same vertical and horizontal chains of data qubits as in Fig.~\ref{fig:surface_code_geometry}, with the Pauli type alternating between $X$ and $Z$ along each chain.
We denote these mixed-type logical operators as $L_V$ and $L_H$, which play the same roles as the logical $X$ and $Z$ observables of the rotated surface code.
The ordering of CZ gates applied between each data qubit and the ancilla qubit at the centre of a face follows an N-shape, depicted in Fig.~\ref{fig:se_circuit}(c).

\begin{figure}[t]
  \centering
  \begin{subfigure}[b]{\columnwidth}
  \includegraphics[width=\columnwidth]{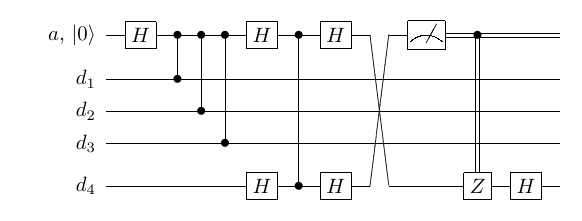}
  \caption{Z stabiliser}
  \end{subfigure}
  \begin{subfigure}[b]{\columnwidth}
  \includegraphics[width=\columnwidth]{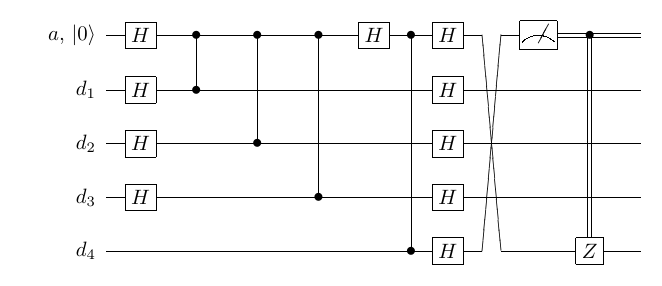}
  \caption{X stabiliser}
  \end{subfigure}
  \caption{SWAP SE circuits for the surface code. 
  This circuit combines the normal SE circuit with a qubit-state SWAP gate and a physical exchange of atom positions. CZ gate cancellations in the SWAP gate decomposition allow SWAP SE to be implemented with no additional entangling gates compared with normal SE. 
  This procedure is intended to mitigate atom loss and heating on data qubits~\cite{suchara2015leakage, Baranes2026}.}
  \label{fig:swap_se}
\end{figure}

Depending on the ordering of two-qubit gate operations, along with atom transportation, several variants of syndrome extraction are possible:

\paragraph{Normal SE.}
One variant is standard syndrome extraction, referred to as normal SE.
In this method, two-qubit gates are applied to all ancilla qubits in a uniform ordering, followed by direct measurement of the ancilla qubits.
To minimise the effect of hook errors, we employ the Z/N ordering, where the application order of CZ gates differs between $X$-type and $Z$-type stabilisers~\cite{O'Rourke2025}.

\paragraph{SWAP SE.}

Another approach is syndrome extraction with SWAP operations, referred to as SWAP SE~\cite{suchara2015leakage, PRXQuantum.5.040343, Perrin2025quantumerror, Baranes2026}. Since qubits suffer from atom loss, such errors can be mitigated by periodically replacing the data qubits with fresh atoms. SWAP SE provides a mechanism for this by exchanging the roles and positions of ancilla and data qubits during syndrome extraction. 

In neutral-atom systems, SWAP SE can be implemented by combining a qubit-state SWAP gate with physical atom transport. The state SWAP exchanges the logical quantum states of ancilla and data qubits, while the physical SWAP exchanges their positions using movable optical tweezers. The qubit-state SWAP gate can be decomposed into three CZ gates together with Hadamard gates. This SE circuit can be naturally implemented in single-isotope systems, where data and ancilla qubits have the same atomic properties and can therefore be interchanged without changing the control operations. 

Importantly, SWAP SE can be implemented with the same number of entangling gates as normal SE as shown in Fig.~\ref{fig:swap_se}. Although a SWAP gate requires three CZ gates, the final CZ gate of the normal stabiliser measurement cancels with the first CZ gate in the decomposed SWAP gate. The final CZ gate of the SWAP decomposition can also be absorbed into the final measurement as a classically controlled $Z$ gate on the data qubit controlled by the ancilla qubit. 

This approach mitigates the accumulation of lost atoms on data qubits by replacing them with fresh ancilla atoms. However, it simultaneously introduces other sources of noise compared to normal SE. Moreover, SWAP SE cannot be performed in dual-isotope systems with in-place measurement, because the data and ancilla qubits are different isotopes and so cannot be interchanged directly. We therefore use SWAP SE only for single-isotope configurations.

\begin{figure}[t]
  \centering
  \includegraphics[width=\columnwidth]{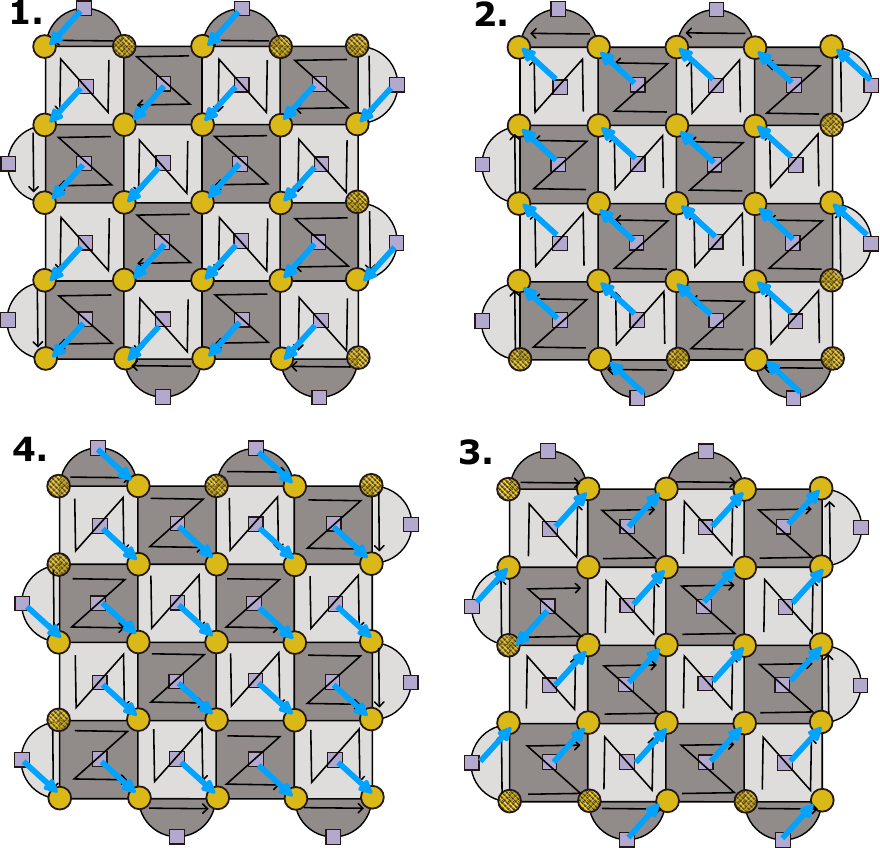}
  \caption{Dynamical-ZN ordering for SWAP SE. The CZ gate order (black N/Z-shape arrows) differs between $X$-type and $Z$-type stabilisers~\cite{O'Rourke2025}. The blue arrow represents the replacement of data qubits by the ancilla qubit. The replaced qubit changes at each stabiliser round to maximise the number of data qubits refreshed as ancilla qubits via SWAP operations. The cross-hatched data qubits are not refreshed in each respective round.}
  \label{fig:swap_se_order}
\end{figure}

For SWAP SE, certain data qubits at the boundary of the array remain un-swapped when using a fixed syndrome readout order.
To maximise the number of data qubits refreshed via SWAP operations, we also consider a \emph{dynamical-ZN} ordering as shown in Fig.~\ref{fig:swap_se_order}, in which the gate ordering alternates between consecutive stabiliser rounds.
We hereafter refer to this SE variant, i.e. with SWAP operations and dynamical ordering, as SWAP SE.

\paragraph{Simulation setup.}
In this work, we simulate memory experiments of the rotated and XZZX surface codes.
For dual-isotope systems, we assume normal SE with in-place measurements, while we assume several different configurations for single-isotope systems. There are two options for the measurement method in single-isotope systems, that is, shelving and zoned measurement. In addition, there are two options for the syndrome extraction method in single-isotope systems, that is, normal SE and SWAP SE. We consider all configurations of these options.

The Clifford circuit simulation is performed via Monte Carlo sampling using \texttt{stim}~\cite{gidney2021stim}.
Noisy gates are inserted into the \texttt{stim} circuits using \texttt{DualYbSim}, which is a Python library developed for simulating noise channels of dual-Yb systems as a wrapper of stim circuits~\cite{DualYbSim}.
For each data point, we sample $10^5$--$10^7$ shots to estimate the LER and its variance.
To decode syndromes, we use \texttt{PyMatching}~\cite{Higgott2025sparseblossom}, a standard decoder package for minimum-weight perfect matching decoders for the surface code.
We utilise the correlated matching option~\cite{Fowler2013} provided by pymatching to decode syndromes.

The noise channels and parameters we assume are based on current experimental demonstrations and expectations for their improvement. The details of the approximated noise channels are described and justified in Appendix~\ref{app:noise_channel_definitions}, and the parameters are given with relevant references in Table~\ref{tab:noise_channel_summary}.

\section{Results}
\label{sec:results}

\subsection{Logical error rates for each isotope configuration}
\label{subsec:ler}

\quad To evaluate the performance of the surface codes on Yb systems, we compute the LER per round $p_L$ for each configuration of isotopes, encoding, and SE method.
Setting $r$ to be the number of SE rounds and $p_{L,r}$ to be the LER after $r$ rounds, then the LER per round $p_L$ is defined as
\begin{equation}
  p_L = 1 - (1 - p_{L,r})^{1/r} \simeq \frac{p_{L,r}}{r}.
  \label{eq:ler_per_round}
\end{equation}
 $p_L \ll 1$ is assumed here for the approximation, which holds for the parameter region we are interested in. Hence, we obtain $p_L$ from the right hand side of Eq.~\eqref{eq:ler_per_round} hereafter.
The number of SE rounds is set to be $r = d$ for every configuration in this work, where $d$ is the code distance.

We employed normal SE for the dual-isotope cases, while both normal SE and SWAP SE were employed for single-isotope cases. As mentioned in Section~\ref{sec:surface_code}, we also consider three types of measurement methods, that is, in-place measurement, shelving, and zoned measurement.

\begin{figure}
  \centering
  \includegraphics[width=\columnwidth]{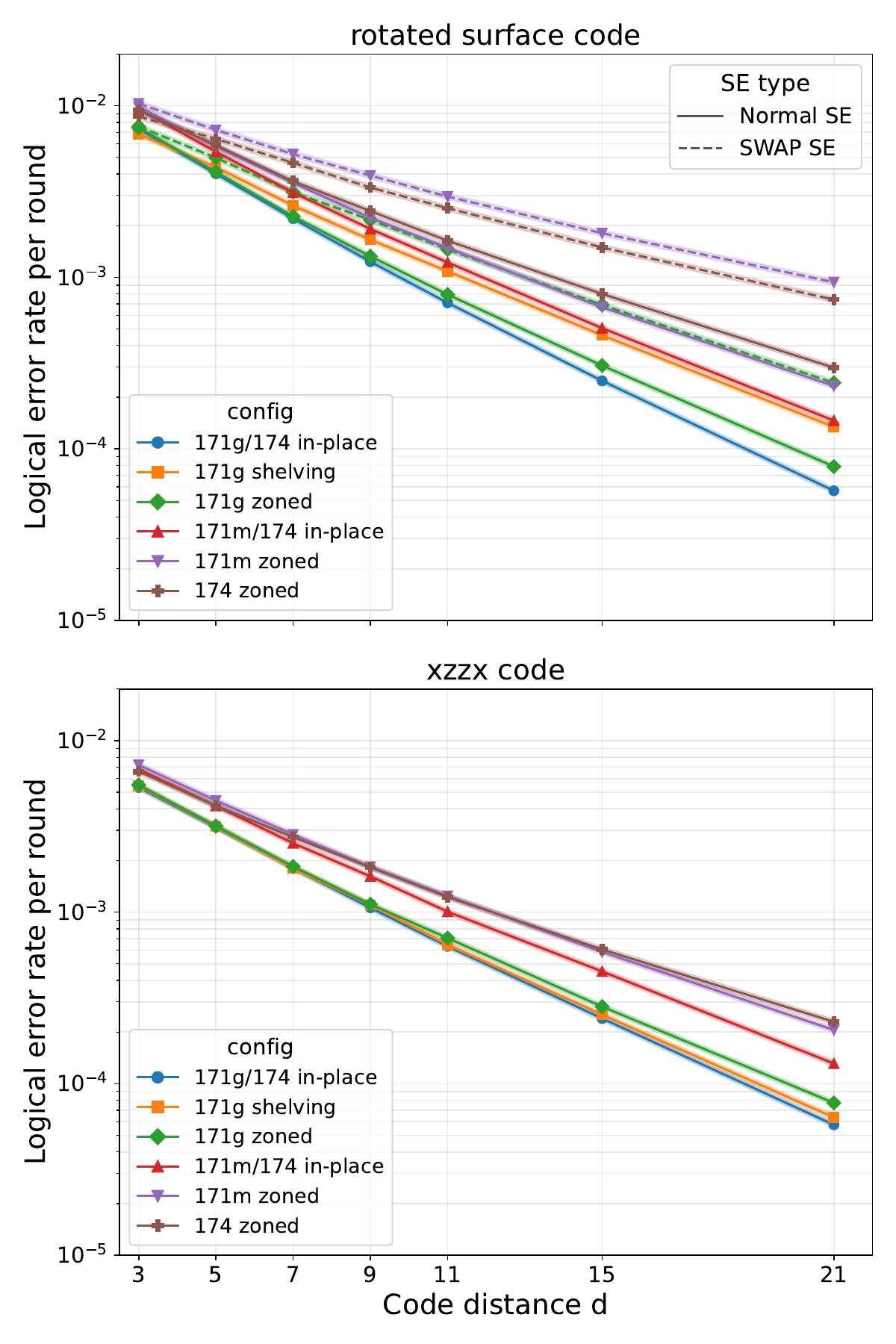}
  \caption{The dependence of the LER per round of the rotated and XZZX surface codes at code distance~$d$ for different configurations of isotope, encoding and SE measurement method.
  Colour and marker delineate the isotope combination and encoding, and the line style represents the SE method. 
  The dual-isotope configuration using normal SE with in-place measurement (blue solid line with circle marker) achieves the lowest LER among all the configurations. The SWAP SE cases (dashed line) achieve similar or higher LER compared to normal SE for all configurations on the rotated surface code.
  Moreover, comparing the two codes, the XZZX code achieves a LER comparable to or lower than that of the rotated surface code.}
  \label{fig:ler_comparison}
\end{figure}

The numerical results are shown in Fig.~\ref{fig:ler_comparison}.
While $p_L$ decreases with increasing code distance for all configurations, implying subthreshold behaviour at the error rates of the model, the gradient of $p_L$ differs across configurations.
As shown in Fig.~\ref{fig:ler_comparison}, the ($\fermi$-g, $\boson$) dual-isotope system achieves the greatest improvement of $p_L$ among configurations for both the XZZX and rotated surface codes.

Comparing the two codes, the XZZX code achieves a $p_L$ comparable to, or lower than, that of the rotated surface code for all isotope configurations.
For the ($\fermi$-g, $\boson$) dual-isotope configurations, the two codes perform almost identically, whereas the XZZX code achieves a lower $p_L$ for single-isotope configurations; for example, in the case of $\fermi$-g qubits with shelving measurement, the $p_L$ is approximately half in XZZX compared to the rotated surface code at $d = 21$.

Considering just the choice of isotope and encoding for the data qubit: $\fermi$-g qubits tend to have better performance than $\fermi$-m qubits.
The principal difference is due to the contribution of decay errors from the metastable state to the ground state, such as $\mathrm{DECAY}_{mg}^{(\mathrm{gate})}$ and $\mathrm{DECAY}_{mg}$ in Appendix~\ref{app:noise_channel_definitions}.
While these decay errors are expected to be detectable by additional efforts, such as erasure conversion~\cite{Wu2022-uy, Sahay2023-zo}, these leakage-detection gadgets introduce additional overhead during SE, necessitating further comparison which we leave for future work.

Additionally, among the single-isotope cases, normal SE yields a lower $p_L$ compared to SWAP SE.
SWAP SE was expected to outperform normal SE from previous studies~\cite{PRXQuantum.5.040343, Perrin2025quantumerror, Baranes2026}, but our numerical results suggest the opposite for Yb systems.
This is because SWAP SE requires additional gate operations and these induce additional errors that are expected to outweigh the benefit of refreshing the atom qubits through SWAP operations.

These results suggest that, for Yb systems, in-place measurement is expected to outperform SWAP SE for syndrome extraction when the erasure information is not being utilised for decoding.
Although the numerical results of Fig.~\ref{fig:ler_comparison} show a disadvantage of other measurement and SE methods, it is expected that the detected erasure information would improve the decoding performance of the surface codes, as previous works have shown~\cite{Perrin2025quantumerror, Baranes2026}.

\subsection{Error budget of dual-Yb systems for the surface code}
\label{subsec:error_budget}

\quad To understand the predominant contributions to logical performance between configurations, we further decompose the inverse of the LER suppression factor into its constituent components, thereby estimating the contribution of each error source and deriving an error budget~\cite{Acharya2021, GoogleQuantumAI2023}.

Letting $p_{L}(d)$ be $p_L$ of a QEC code, with dependence on code distance~$d$ explicitly written, below threshold it is expected to satisfy
\begin{equation}
  p_{L}(d) = \frac{C}{\Lambda^{(d+1)/2}},
  \label{eq:ler_scaling}
\end{equation}
where $C$ is a constant and $\Lambda$ is the error suppression factor. 
From this relation, the suppression factor can be obtained as the gradient of the LER on a logarithmic scale with respect to the code distance, e.g. $\Lambda = {p_{L}(d)}/{p_{L}(d+2)}$.
Letting $p$ and $p_\mathrm{th}$ denote the physical error rate and the QEC threshold respectively, we have $\Lambda^{-1} \propto p / p_\mathrm{th}$.   
Since $\Lambda^{-1}$ is expected to be first order in $p$, it is expected to be well-approximated as a linear function of disparate error rate sources $p_i\in P$, where $P$ and $p_i$ are the set of physical error rates and the physical error rate of each error source respectively.
As long as correlations among physical errors are negligible, 
$\Lambda^{-1}$ can be expressed approximately as a combination of physical error rates $p_i \in P$~\cite{Acharya2021, GoogleQuantumAI2023}:
\begin{equation}
  \Lambda^{-1} \simeq \sum_i \left(w_i \, p_i + h_i p^2_i\right),
  \label{eq:lambda_decomposition}
\end{equation}
where $w_i$ and $h_i$ are the linear and quadratic sensitivity of each $p_i$ independently.
The term $C_i=w_i p_i + h_i p^2_i$ represents the contribution of each $p_i$ to the LER.
While each contribution $C_i$ is estimated from the effective linear sensitivity, which in~\cite{Acharya2021, GoogleQuantumAI2023} is estimated from the gradient of the LER when $p_i\in P$ is half, here we instead take $C_i$ as a quadratic function of $p_i\in P$ that is derived by curve fitting explained in the following paragraph. This omits quadratic cross-terms $p_ip_j$, which are instead incorporated into the $w_ip_i$ terms here.
A smaller value of $\Lambda^{-1}$ (i.e., a larger $\Lambda$) indicates a greater reduction in LER with increasing code distance, implying more efficient error suppression.
$\Lambda^{-1}$ is not truly linear, hence the inclusion of the quadratic sensitivities that are expected to account for the majority of higher-order contributions and thus it serves as a good approximation. 

We estimate $\Lambda^{-1}$ and the expansion coefficients $w_i$ as follows.
For each error parameter $p_i \in P$, we prepare noise parameter sets where only $p_i$ is scaled by factors from $0.4$ to $1.1$ in increments of $0.025$.
For each parameter set, we estimate the LER $p_{L}(d)$ and its variance $V(p_{L}(d))$ at code distances $d = 5$ and $d = 7$ from either $10^7$ samples taken or $10^5$ error events taken.
From these samples, we estimate $\Lambda^{-1}$ and its variance $V(\Lambda^{-1})$ via error propagation.
Using the inverse of $V(\Lambda^{-1})$ as the weight for each data point, we use a weighted quadratic program to estimate the coefficients $w_i$ and $h_i$. Heuristically, we only estimate the coefficient $h_i$ for errors related to Rydberg decay because the contributions of other quadratic terms are expected to be sufficiently small for $p\ll 1$. Any difference $\Lambda^{-1} - \sum_iC_i$ is taken as \textit{stray error} which represents the contribution that we cannot capture from the quadratic approximation of \eqref{eq:lambda_decomposition}, occurring close to threshold or due to distance-dependent noise sources.
The quadratic programming is solved using \texttt{CVXPY}~\cite{cvxpy}.

We estimate the error budget of the surface code over different configurations of isotope, qubit encoding, and measurement scheme. For the ($\fermi$-g, $\boson$) dual-isotope configuration, we consider in-place measurement. For the $\fermi$-g single-isotope architectures, we consider both shelving and zoned measurement schemes. For the $\boson$ clock qubits, we consider zoned measurement. In each configuration, normal SE is utilised when evaluating the error budget.

\begin{figure}[t]
  \centering
  \includegraphics[width=\columnwidth]{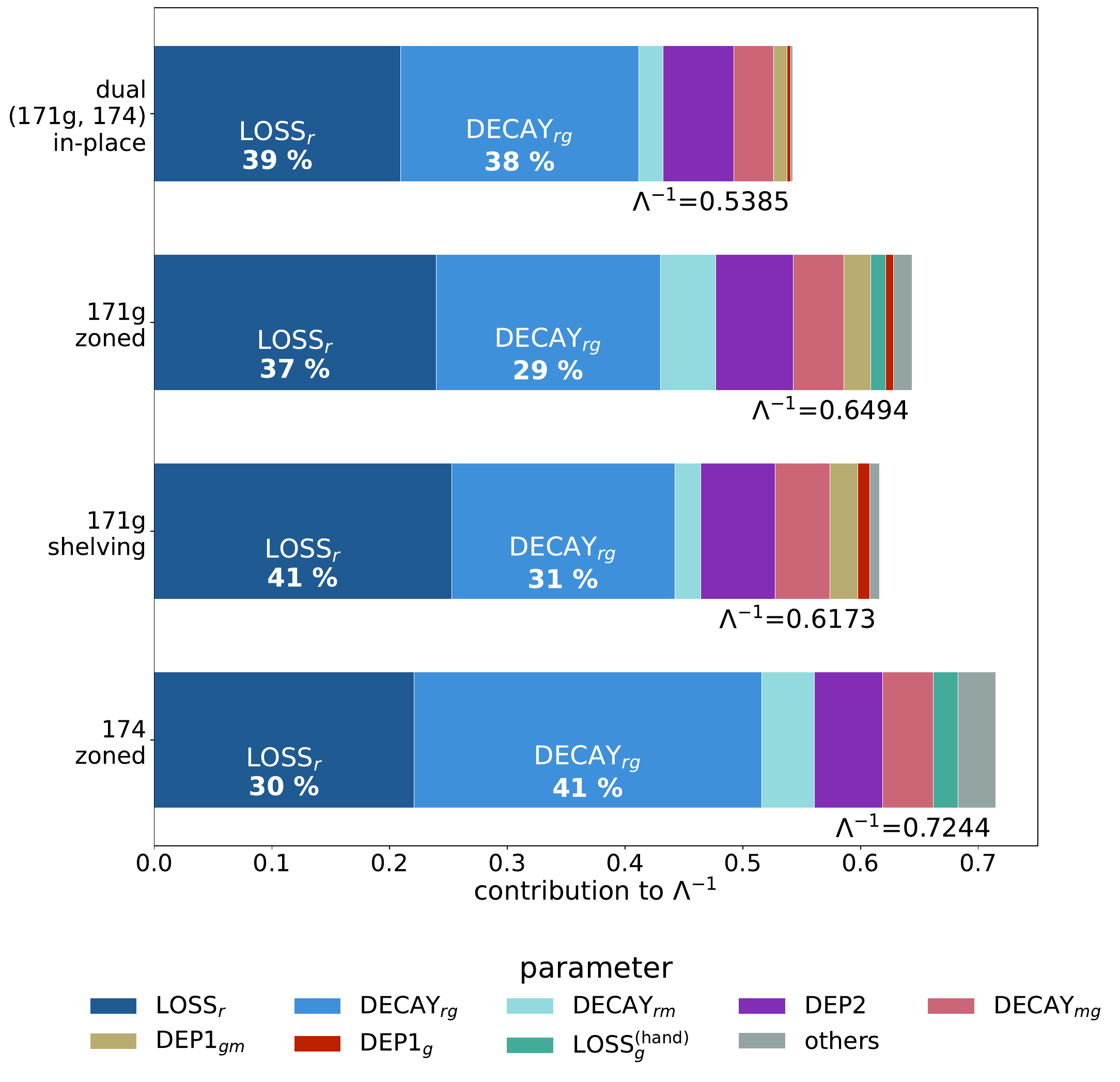}%
  \caption{Error budgets for ($\fermi$-g, $\boson$), $\fermi$-g single-isotope with zoned measurement, $\fermi$-g single-isotope with shelving, and $\boson$ clock qubits with zoned measurement. This plot shows the division of the error contribution of each $p_i\in P$ to $\Lambda^{-1}$ as $C_i$. The main contributions are the decay errors from the Rydberg state during the application of two-qubit gates. The next most significant contributions are from two-qubit depolarising noise on two-qubit gates and metastable-to-ground decay. Small contributions and stray error that cannot be captured by the linear approximation are attributed to ``others''.}
  \label{fig:error_budget}
\end{figure}

The numerical results are shown in Fig.~\ref{fig:error_budget}.
$\Lambda^{-1}$ of the dual-isotope architecture is approximately 0.54 while the $\Lambda^{-1}$ of the other configurations ranges approximately from 0.62 to 0.72.
This shows the dual-isotope configuration would be a better candidate for implementing a surface code memory.
In all cases, decay from the Rydberg state is the dominant noise source. The contribution fraction in each case is 80~\% for the ($\fermi$-g, $\boson$) dual-isotope configuration, 74~\% for the $\fermi$-g qubits with zoned measurement, 75~\% for the $\fermi$-g qubits with shelving measurement, and 78~\% for the $\boson$ clock qubits with zoned measurement. 
This analysis demonstrates that improving the quality of the two-qubit gate, particularly due to Rydberg decay, is the most promising pathway to achieving a higher error suppression factor.

The sensitivity weights $w_i$ obtained during the contribution analysis are listed in Table~\ref{tab:gamma-rl-branched-error-budget} for the ($\fermi$-g, $\boson$) configuration. Here, low contributions $|C_i|\le 10^{-10}$ are omitted.
For the ($\fermi$-g, $\boson$) configuration, $\Lambda$ is particularly sensitive to depolarising errors during clock excitation performed for two-qubit gates.
A similar sensitivity to depolarising errors is observed for configurations of $\fermi$-g single-isotope architectures, which also rely on clock excitation for two-qubit gates.

Depolarising error parameters such as $p_1^{(g)}$, $p_1^{(gm)}$, and $p_1^{(c)}$ are expected to improve with advances in control techniques, so their contributions may be reduced with such advances.
In contrast, errors originating from atomic decay rates reflect intrinsic properties of the atoms that are difficult to reduce directly.
However, in the regime where the gate time $t$ is sufficiently short, these contributions scale linearly as $w_i p_i \simeq w_i \Gamma t$, suggesting that reducing the gate times may also mitigate their impact.
For example, in the ($\fermi$-g, $\boson$) configuration, suppressing the Rydberg decay contribution to a level comparable to other error sources (${\sim}0.02$) would require reducing the gate time by a factor of ${\sim}1/20$ (to ${\sim}1.5 \times 10^{-8}$~s).
Such improvements may be achievable with higher-power pulsed lasers and related technical advances~\cite{Chew2022Ultra, magro2026}.

\begin{figure}
  \centering
  \includegraphics[width=\columnwidth]{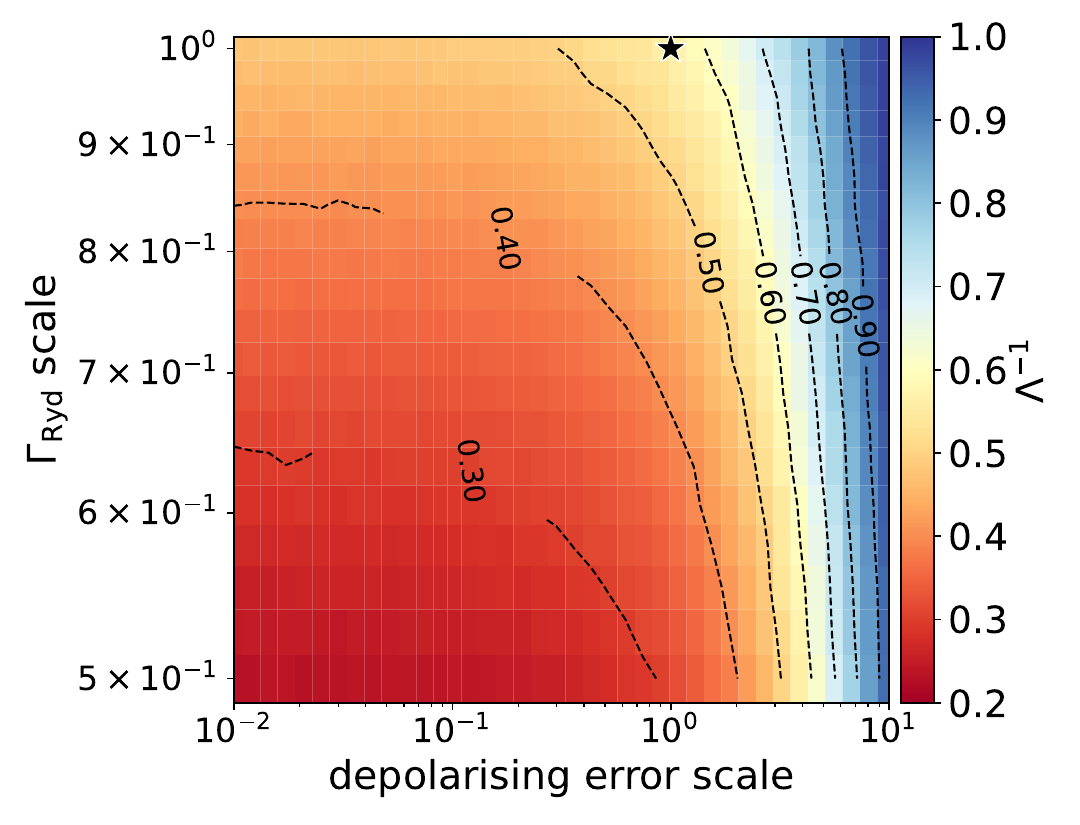}%
  \caption{Error suppression factor for the ($\fermi$-g, $\boson$) configuration when changing the Rydberg decay rate and depolarising error rate by a scaling factor. The baseline error parameters are set to the values shown in Table~\ref{tab:gamma-rl-branched-error-budget}, indicated as a black star. Each error rate is swept over a multiplicative factor, from $10^{-2}$ to $10$ for the depolarising error rates and from $0.5$ to $1$ for the Rydberg decay rate. This plot suggests that improvement of the decay rate from the Rydberg state is more effective than comparable improvements in the depolarising error rate.}
  \label{fig:heatmap_lambda_inv_dual}
\end{figure}

We have also evaluated how the error suppression factor changes if we were able to reduce either the Rydberg decay rate or the depolarising error rate, as shown in Fig.~\ref{fig:heatmap_lambda_inv_dual}.
This plot suggests that $\Lambda^{-1}$ can be significantly improved by small improvements in both depolarising errors on each gate operation and the Rydberg decay $\Gamma_{\textrm{Ryd}}t$, where $\Gamma_{\textrm{Ryd}}$ denotes the radiative decay rate of the Rydberg state. 
We can also see that efforts in improving the decay error rate to half is more effective than similar efforts towards depolarising error rates.
For example, once $\Gamma_\mathrm{Ryd}$ is reduced to half, $\Lambda^{-1}\simeq\;0.54\to\;0.32$, meaning the LER gets suppressed 1.7 times faster.
Heatmaps for other configurations can be found in Fig.~\ref{fig:app:heatmap_lambda_inv_single_isotope} included in Appendix~\ref{app:effects_of_error_reduction_for_Lambda}.

\begin{table*}[!tp]
\caption{Contributions to the error budget of Fig.~\ref{fig:error_budget}. The error factors with low contributions ($|C_i|\le 10^{-10}$) are omitted in this table. Here $p_i$ denotes the baseline error strength used in the fit. For channels associated with Rydberg-state decay, $h_i$ denotes the quadratic sensitivity; other parameters are linear-only. Noise-channel definitions are given in Table~\ref{tab:noise_channel_summary} and Appendix~\ref{app:noise_channel_definitions}.}
  \label{tab:gamma-rl-branched-error-budget}
  \centering
  \scriptsize
  \begin{tabular}{llrrrr}
  \hline
  Case & Noise channel & Strength $p_i$ & Contribution $C_i$ & sensitivity $w_i$ & Quad. sensitivity $h_i$ \\
  \hline
  171g/174 in-place & $\mathrm{LOSS}_{r}$ ($\fermi, \boson$) & $3.055\times 10^{-3}$ & $2.092\times 10^{-1}$ & $6.333\times 10^{1}$ & $1.685\times 10^{3}$ \\
   & $\mathrm{DECAY}_{rg}$ ($\fermi, \boson$) & $2.517\times 10^{-3}$ & $2.023\times 10^{-1}$ & $8.336\times 10^{1}$ & $-1.183\times 10^{3}$ \\
   & $\mathrm{DEP2}_{\mathrm{dual}}$ ($\fermi, \boson$) & $1.000\times 10^{-3}$ & $6.027\times 10^{-2}$ & $6.027\times 10^{1}$ & -- \\
   & $\mathrm{DECAY}_{rm}$ ($\fermi, \boson$) & $4.199\times 10^{-4}$ & $2.061\times 10^{-2}$ & $0$ & $1.169\times 10^{5}$ \\
   & $\mathrm{DECAY}_{mg}$ ($\fermi$) & $9.995\times 10^{-4}$ & $1.915\times 10^{-2}$ & $1.916\times 10^{1}$ & -- \\
   & $\mathrm{DECAY}_{mg}$ ($\boson$) & $9.995\times 10^{-4}$ & $1.443\times 10^{-2}$ & $1.444\times 10^{1}$ & -- \\
   & $\mathrm{DEP1}_{gm}$ & $1.000\times 10^{-4}$ & $1.162\times 10^{-2}$ & $1.162\times 10^{2}$ & -- \\
   & $\mathrm{DEP1}_{g}$ & $1.000\times 10^{-4}$ & $2.907\times 10^{-3}$ & $2.907\times 10^{1}$ & -- \\
   & $\mathrm{LOSS}_{g}^{(\mathrm{reset})}$ ($\boson$) & $1.000\times 10^{-3}$ & $7.332\times 10^{-4}$ & $7.332\times 10^{-1}$ & -- \\
   & $\mathrm{DEP1}_{c}$ & $1.000\times 10^{-4}$ & $6.826\times 10^{-4}$ & $6.826\times 10^{0}$ & -- \\
   & $\mathrm{LOSS}_{g}^{(\mathrm{meas})}$ ($\boson$) & $1.000\times 10^{-3}$ & $3.626\times 10^{-4}$ & $3.626\times 10^{-1}$ & -- \\
   & $\mathrm{ZERR}_{c}$ & $2.000\times 10^{-5}$ & $4.790\times 10^{-5}$ & $2.395\times 10^{0}$ & -- \\
  \hline
  171g zoned & $\mathrm{LOSS}_{r}$ & $3.055\times 10^{-3}$ & $2.395\times 10^{-1}$ & $7.424\times 10^{1}$ & $1.358\times 10^{3}$ \\
   & $\mathrm{DECAY}_{rg}$ & $2.517\times 10^{-3}$ & $1.904\times 10^{-1}$ & $6.659\times 10^{1}$ & $3.603\times 10^{3}$ \\
   & $\mathrm{DEP2}_{m}$ & $1.000\times 10^{-3}$ & $6.590\times 10^{-2}$ & $6.590\times 10^{1}$ & -- \\
   & $\mathrm{DECAY}_{rm}$ & $4.199\times 10^{-4}$ & $4.699\times 10^{-2}$ & $1.631\times 10^{2}$ & $-1.220\times 10^{5}$ \\
   & $\mathrm{DECAY}_{mg}$ & $9.995\times 10^{-4}$ & $4.313\times 10^{-2}$ & $4.315\times 10^{1}$ & -- \\
   & $\mathrm{DEP1}_{gm}$ & $1.000\times 10^{-4}$ & $2.256\times 10^{-2}$ & $2.256\times 10^{2}$ & -- \\
   & $\mathrm{LOSS}_{g}^{(\mathrm{hand})}$ & $1.000\times 10^{-3}$ & $1.297\times 10^{-2}$ & $1.297\times 10^{1}$ & -- \\
   & $\mathrm{FLIP}_{g}$ (measurement, reset) & $1.000\times 10^{-3}$ & $7.926\times 10^{-3}$ & $7.926\times 10^{0}$ & -- \\
   & $\mathrm{DEP1}_{g}$ & $1.000\times 10^{-4}$ & $6.296\times 10^{-3}$ & $6.296\times 10^{1}$ & -- \\
   & $\mathrm{ZERR}_{g}$ & $1.000\times 10^{-4}$ & $6.209\times 10^{-3}$ & $6.210\times 10^{1}$ & -- \\
   & $\mathrm{MERR}$ & $1.000\times 10^{-4}$ & $1.092\times 10^{-3}$ & $1.092\times 10^{1}$ & -- \\
   & $\mathrm{LOSS}_{g}^{(\mathrm{meas})}$ & $1.000\times 10^{-3}$ & $9.267\times 10^{-4}$ & $9.267\times 10^{-1}$ & -- \\
  \hline
  171g shelving & $\mathrm{LOSS}_{r}$ & $3.055\times 10^{-3}$ & $2.526\times 10^{-1}$ & $8.559\times 10^{1}$ & $-9.547\times 10^{2}$ \\
   & $\mathrm{DECAY}_{rg}$ & $2.517\times 10^{-3}$ & $1.896\times 10^{-1}$ & $6.557\times 10^{1}$ & $3.887\times 10^{3}$ \\
   & $\mathrm{DEP2}_{m}$ & $1.000\times 10^{-3}$ & $6.289\times 10^{-2}$ & $6.289\times 10^{1}$ & -- \\
   & $\mathrm{DECAY}_{mg}$ & $9.995\times 10^{-4}$ & $4.663\times 10^{-2}$ & $4.666\times 10^{1}$ & -- \\
   & $\mathrm{DEP1}_{gm}$ & $1.000\times 10^{-4}$ & $2.377\times 10^{-2}$ & $2.377\times 10^{2}$ & -- \\
   & $\mathrm{DECAY}_{rm}$ & $4.199\times 10^{-4}$ & $2.213\times 10^{-2}$ & $0$ & $1.255\times 10^{5}$ \\
   & $\mathrm{DEP1}_{g}$ & $1.000\times 10^{-4}$ & $1.026\times 10^{-2}$ & $1.026\times 10^{2}$ & -- \\
   & $\mathrm{FLIP}_{g}$ (measurement, reset) & $1.000\times 10^{-3}$ & $7.091\times 10^{-3}$ & $7.091\times 10^{0}$ & -- \\
   & $\mathrm{LOSS}_{g}^{(\mathrm{meas})}$ & $1.000\times 10^{-3}$ & $1.002\times 10^{-3}$ & $1.002\times 10^{0}$ & -- \\
  \hline
  174 zoned & $\mathrm{DECAY}_{rg}$ & $2.517\times 10^{-3}$ & $2.953\times 10^{-1}$ & $1.407\times 10^{2}$ & $-9.307\times 10^{3}$ \\
   & $\mathrm{LOSS}_{r}$ & $3.055\times 10^{-3}$ & $2.207\times 10^{-1}$ & $7.691\times 10^{1}$ & $-1.536\times 10^{3}$ \\
   & $\mathrm{DEP2}_{c}$ & $1.000\times 10^{-3}$ & $5.781\times 10^{-2}$ & $5.781\times 10^{1}$ & -- \\
   & $\mathrm{DECAY}_{rm}$ & $4.199\times 10^{-4}$ & $4.490\times 10^{-2}$ & $1.514\times 10^{2}$ & $-1.059\times 10^{5}$ \\
   & $\mathrm{DECAY}_{mg}$ & $9.995\times 10^{-4}$ & $4.330\times 10^{-2}$ & $4.333\times 10^{1}$ & -- \\
   & $\mathrm{LOSS}_{g}^{(\mathrm{hand})}$ & $1.000\times 10^{-3}$ & $2.106\times 10^{-2}$ & $2.106\times 10^{1}$ & -- \\
   & $\mathrm{DEP1}_{c}$ & $1.000\times 10^{-4}$ & $9.644\times 10^{-3}$ & $9.644\times 10^{1}$ & -- \\
   & $\mathrm{ZERR}_{c}$ & $2.000\times 10^{-5}$ & $8.203\times 10^{-3}$ & $4.101\times 10^{2}$ & -- \\
   & $\mathrm{LOSS}_g^{(\mathrm{reset})}$ & $1.000\times 10^{-3}$ & $6.756\times 10^{-3}$ & $6.756\times 10^{0}$ & -- \\
   & $\mathrm{MERR}$ & $1.000\times 10^{-4}$ & $3.667\times 10^{-3}$ & $3.667\times 10^{1}$ & -- \\
   & $\mathrm{LOSS}_{g}^{(\mathrm{meas})}$ & $1.000\times 10^{-3}$ & $2.287\times 10^{-3}$ & $2.287\times 10^{0}$ & -- \\
  \hline
  \end{tabular}
  \end{table*}

\section{Conclusion}
\label{sec:conclusion}

\quad We have numerically evaluated and compared the performance of surface-code memories for dual-isotope and single-isotope Yb configurations, closely taking into account realistic sources of noise with experimentally motivated parameters.
Our results show that the dual-isotope configuration, using the $\fermi$ ground-state qubit as the data qubit and the $\boson$ optical clock-state qubit as the ancilla with in-place measurement, achieves the best performance in terms of LER. This advantage arises because the dual-isotope configuration avoids additional recurring errors associated with shelving and zoned measurement, which are required for syndrome extraction in single-isotope configurations. 

Our error-budget analysis clearly shows that the dominant error source is Rydberg decay during two-qubit gate operations. This highlights the importance of fundamental research aimed at reducing decay errors during CZ gates. Reducing Rydberg decay errors would improve the surface-code memory performance for all configurations. However, if the Rydberg decay contribution is completely omitted from the error budget, the remaining contributions are lowest for the dual-isotope configuration. This suggests that even as the impact of Rydberg decay is reduced, this configuration will still have the best performance. If Rydberg decay errors remain a limiting factor, loss-aware decoding techniques~\cite{Perrin2025quantumerror,Baranes2026,liu2026Loss,nishio2026,pavlovich2026} would play an important role in improving the surface-code LER. While atom reloading is beyond the scope of the present work, efficient reloading strategies for correcting atom loss~\cite{Chiu2025Continuous, Li2025Fast, Kobayashi2026, zhang2025lev, reichardt2025} and careful noise modelling of the associated processes would be valuable extensions of such loss-aware approaches.

Our results provide quantitative insight into the design of near-term neutral-atom quantum computers. This evaluation can be extended to other quantum error-correcting codes, including high-rate QEC codes~\cite{Bravyi2024BB,Goto2024,Yoshida2025C4C6,Xu2024Constant,tamiya2026fault,zhao2026ultra}, and to other neutral-atom species configurations, such as rubidium and caesium~\cite{anand2024dual,miles2026}. Various research directions in decoding will be significant in this broader context, whereby further reductions in LER could be obtained by incorporating soft information such as erasure flags~\cite{gottesman1997,Wu2022-uy,Ma2023high,Scholl2023-gw, PRXQuantum.5.040343} or complementary gap data~\cite{YokedSurfaceCode2025}, or by employing AI-based decoders~\cite{Bausch2024,Senior2026}.
 
\begin{acknowledgments}
\quad We thank the members of Yaqumo Inc. for extensive discussions on the experimental aspects of neutral-atom quantum computers. We especially thank Sylvain de L\'{e}s\'{e}leuc and Takafumi Tomita for carefully reading the manuscript and providing valuable feedback. 
F.K. was supported by JSPS KAKENHI Grant Number 25KJ0445. T.K. was supported by JST SPRING Grant No. JPMJSP2110. N.F. and Y.N. were supported by Project No. JPNP25014, funded by the New Energy and Industrial Technology Development Organization (NEDO), Japan. 

FK performed the numerical simulations and devised the Pauli-twirling approximation, SE strategy, LER analysis, and error-budget analysis. TK developed the Yb-specific noise model. NF and YN supervised the project and contributed to developing and validating the theoretical framework and its physical justification. FK, TK, and NF wrote the manuscript. All authors discussed and interpreted the results and reviewed the manuscript.
We acknowledge the use of AI to assist in the coding of numerical simulations, figure production, and rigorous grammar checking for the final manuscript.
\end{acknowledgments}

\section*{Data Availability}
The data supporting the findings of this study can be made available from the authors upon reasonable request.

\appendix
\section*{Appendix}
\setcounter{table}{0}
\renewcommand{\thetable}{A\arabic{table}}
\setcounter{figure}{0}
\renewcommand{\thefigure}{A\arabic{figure}}

\section{Noise-channel definitions}
\label{app:noise_channel_definitions}
\quad In order to evaluate the memory experiments considered in this work, we developed a numerical noise model that is detailed throughout this appendix. We describe each noise process first as a set of Kraus operators, which are subsequently mapped onto an effective Pauli and erasure error model using generalised Pauli twirling.

For reference, Table~\ref{tab:noise_channel_summary} summarises all noise channels included in the simulations, together with the applicable isotopes, activation conditions, and experimental parameters. The table serves as an index; complete channel definitions with Kraus representations are provided below.

This appendix is organised as follows. Section~\ref{app:noise_conventions} states the notation and channel conventions. Sections~\ref{app:coherent_control_channels}, \ref{app:measurement_channels}, \ref{app:reset_channels}, \ref{app:idling_channels}, \ref{app:decay_channels}, and \ref{app:transportation_channels} provide the Kraus representations for coherent-control, measurement, reset, idling, decay, and transportation channels, respectively. Section~\ref{app:gate_idling_time} summarises the operation times used to set time-dependent error probabilities. Section~\ref{app:channel_ordering} specifies the channel ordering used when multiple noise channels are applied within a circuit operation, and Sec.~\ref{app:channel_ordering_verification} describes its justification.

\begin{figure*}
  \centering
  \begin{subfigure}{0.49\linewidth}
    \centering
    \includegraphics[width=\linewidth]{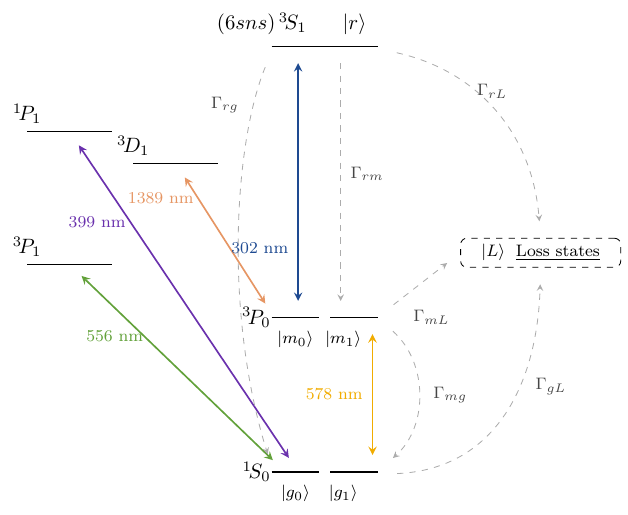}
    \caption{$\fermi$}
    \label{fig:171Yb_diagram}
  \end{subfigure}
  \hfill
  \begin{subfigure}{0.49\linewidth}
    \centering
    \includegraphics[width=\linewidth]{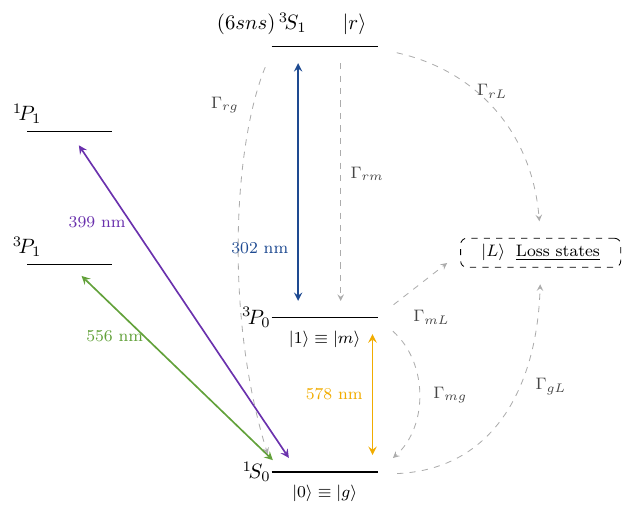}
    \caption{$\boson$}
    \label{fig:174Yb_diagram}
  \end{subfigure}
  \caption{Relevant energy levels diagrams for $\fermi$ and $\boson$.
  (a) $\fermi$ has a nuclear spin $I=1/2$, so that both the $\sSz$ ground state and the $\tPz$ metastable state consist of two nuclear-spin states. $\fermi$ qubits are encoded into either manifold: the ground-state nuclear-spin ($\fermi$-g qubit, $\{\ket{g_0},\ket{g_1}\}$) or the metastable-state ($\fermi$-m qubit, $\{\ket{m_0},\ket{m_1}\}$).
  (b) $\boson$ has zero nuclear spin and hence no hyperfine structure. So $\boson$ qubits are encoded as optical clock qubits, $\ket{0}\equiv\ket{g}$ ($\sSz$) and $\ket{1}\equiv\ket{m}$ ($\tPz$).
  Coloured solid arrows (purple, green, yellow and blue) denote laser-driven transitions, the 399~nm $\sSz\leftrightarrow\sPo$ and 556~nm $\sSz\leftrightarrow\tPo$ transitions are used for ground-state fluorescence imaging and cooling, the 578~nm $\sSz\leftrightarrow\tPz$ clock transition is used for coherent transfer between the ground and metastable manifolds (state-selective readout, shelving, and Rydberg excitation of the $\fermi$-g qubit), and the 302~nm single-photon excitation from $\tPz$ to the Rydberg state $\ket{r}$ of the $(6sns)\,\tSo$ manifold mediates the two-qubit gate. 
  For $\fermi$, the 1389~nm $\tPz\leftrightarrow\tDo$ transition (orange) is used for single-qubit operations within the metastable manifold.
  This operation induces photon scattering from the metastable to the ground-state manifold, which we model as the channel $\mathrm{DECAY}_{mg}^{(\mathrm{gate})}$ during the control of $\fermi$-$m$ qubits (see Appendix~\ref{app:coherent_control_channels}). 
  Grey dashed arrows denote the incoherent decay and loss pathways retained in the noise model, whose rates $\Gamma_{rg}$, $\Gamma_{rm}$, $\Gamma_{rL}$, $\Gamma_{mg}$, $\Gamma_{gL}$, and $\Gamma_{mL}$ are listed in Table~\ref{tab:noise_channel_summary}. The dashed box labelled $\ket{L}$ is the effective loss state, which collects all states outside the retained ground, metastable, and Rydberg manifolds, including atom loss and dark states such as the $\tPt$ manifold.
  }
  \label{fig:energy-level_diagram}
\end{figure*}

\begin{table*}[!tp]
  \centering
  \caption{Summary of the noise channels used in the numerical model. For channels parameterised by a rate $\Gamma$ (including rates written as $1/T_1$ or $1/T_2$), the error probability over an operation of duration $t$ is $p(t)=1-e^{-\Gamma t}$. The classification labels are defined as follows:
  \emph{Coherent-control}, processes occurring only during coherent-control gate operations;
  \emph{Measurement}, processes associated with state detection;
  \emph{Reset}, processes associated with state preparation or reset;
  \emph{Idling}, processes occurring while a qubit is waiting and not actively manipulated, which are disabled during coherent-control operations and reset;
  \emph{Decay}, incoherent population transfer processes from excited states to lower-energy internal states; and
  \emph{Transportation}, processes associated with atom transport and trap handover.
  }
  \label{tab:noise_channel_summary}
  \small
  \setlength{\tabcolsep}{3pt}
  \renewcommand{\arraystretch}{1.08}
  \begin{tabular}{@{}p{0.14\textwidth}p{0.08\textwidth}p{0.14\textwidth}p{0.38\textwidth}p{0.17\textwidth}@{}}
    \toprule
    Noise channel & Isotope & Classification & Value used & References \\
    \midrule
    $\mathrm{DEP1}_{c}$ & 174 & Coherent-control & $p_{1}^{(c)}=0.01\%$ & \cite{Finkelstein2024} \\
    $\mathrm{DEP1}_{gm}$ & 171 & Coherent-control & $p_{1}^{(gm)}=0.01\%$ & \cite{Lis2023,AC2025High,Yan2025} \\
    $\mathrm{DEP1}_{g}$ & 171 & Coherent-control & $p_{1}^{(g)}=0.01\%$ & \cite{Jenkins2022, Ma2022, Lis2023, kusano2025, AC2025High} \\
    $\mathrm{DEP1}_{m}$ & 171 & Coherent-control & $p_{1}^{(m)}=0.01\%$ & \cite{Senoo2025High,Li2025Parallelized,liu2026high} \\
    $\mathrm{DECAY}_{mg}^{(\mathrm{gate})}$ & 171 & Coherent-control & $p_{m\rightarrow g}^{(\mathrm{gate})}=0.1\%$ & \cite{Senoo2025High} \\
    $\mathrm{DEP2}_{c}$ & 174 & Coherent-control & $p_{2}^{(c)}=0.1\%$ & \cite{Tsai2024Bench, Tao2026Sr88, lib2026velocity} \\
    $\mathrm{DEP2}_{m}$ & 171 & Coherent-control & $p_{2}^{(m)}=0.1\%$ & \cite{Senoo2025High, AC2025High, liu2026high} \\
    $\mathrm{DEP2}_{\mathrm{dual}}$ & 171, 174 & Coherent-control & $p_{2}^{(\mathrm{dual})}=0.1\%$ & \cite{white2026, miles2026, wang2026stab} \\
    \midrule
    $\mathrm{LOSS}_{g}^{(\mathrm{meas})}$ & 171, 174 & Measurement & $p_{g\rightarrow L}^{(\mathrm{meas})}=0.1\%$ & \cite{Norcia2023, Lis2023, Huie2023, Muniz2025Repeated, Senoo2025High, zhang2025lev, Falconi2025, yokoyama2026} \\
    $\mathrm{FLIP}_{g}$ & 171 & Measurement & $p_{\mathrm{flip}}^{(g)}=0.1\%$ & \cite{Huie2023, Muniz2025Repeated, Senoo2025High, zhang2025lev} \\
    $\mathrm{MERR}$ & 171, 174 & Measurement & $p_{\mathrm{meas}}=0.01\%$ & \cite{Norcia2023, Lis2023, Huie2023, Muniz2025Repeated, Senoo2025High, zhang2025lev, Falconi2025, yokoyama2026} \\
    \midrule
    $\mathrm{LOSS}_{g}^{(\mathrm{reset})}$ & 171, 174 & Reset & $p_{g\rightarrow L}^{(\mathrm{reset})}=0.1\%$ & \cite{Lis2023,AC2026toric} \\
    $\mathrm{LOSS}_{m}^{(\mathrm{reset})}$ & 171 & Reset & $p_{m\rightarrow L}^{(\mathrm{reset})}=0.6\%$ & \cite{Senoo2025High,zhang2025lev} \\
    $\mathrm{FLIP}_{g}$ & 171 & Reset & $p_{\mathrm{flip}}^{(g)}=0.1\%$ & \cite{Lis2023, reichardt2025} \\
    $\mathrm{FLIP}_{m}$ & 171 & Reset & $p_{\mathrm{flip}}^{(m)}=0.1\%$ & \cite{Senoo2025High} \\
    \midrule
    $\mathrm{ZERR}_{c}$ & 174 & Idling & $1/T_{2}^{(c)}=(5\,\mathrm{s})^{-1}=0.2\,\mathrm{s}^{-1}$ & \cite{young2020half,Finkelstein2024, Ammenwerth2025} \\
    $\mathrm{ZERR}_{gm}$ & 171 & Idling & $1/T_{2}^{(gm)}=(5\,\mathrm{s})^{-1}=0.2\,\mathrm{s}^{-1}$ & \cite{Yan2025, Ma2025Enhancing} \\
    $\mathrm{ZERR}_{g}$ & 171 & Idling & $1/T_{2}^{(g)}=(10\,\mathrm{s})^{-1}=0.1\,\mathrm{s}^{-1}$ & \cite{Jenkins2022, Ma2022, Norcia2023,Lis2023,Huie2023, yuma2024} \\
    $\mathrm{XERR}_{g}$ & 171 & Idling & $1/T_{1}^{(g)}=(200\,\mathrm{s})^{-1}=0.005\,\mathrm{s}^{-1}$ & \cite{Jenkins2022, Ma2022} \\
    $\mathrm{ZERR}_{m}$ & 171 & Idling & $1/T_{2}^{(m)}=(10\,\mathrm{s})^{-1}=0.1\,\mathrm{s}^{-1}$ & \cite{Lis2023, zhang2025lev, Li2025Fast} \\
    $\mathrm{XERR}_{m}$ & 171 & Idling & $1/T_{1}^{(m)}=(200\,\mathrm{s})^{-1}=0.005\,\mathrm{s}^{-1}$ & \cite{zhang2025lev} \\
    \midrule
    $\mathrm{DECAY}_{rg}$ & 171, 174 & Decay & $\Gamma_{rg}=8.4\times10^{3}\,\mathrm{s}^{-1}$\newline $42\%$ branch of a $50\,\mu\mathrm{s}$ Rydberg lifetime & \cite{Wilson2022,Senoo2025High} \\
    $\mathrm{DECAY}_{rm}$ & 171, 174 & Decay & $\Gamma_{rm}=1.4\times10^{3}\,\mathrm{s}^{-1}$\newline $7\%$ branch of a $50\,\mu\mathrm{s}$ Rydberg lifetime & \cite{Wilson2022,Senoo2025High} \\
    $\mathrm{LOSS}_{r}$ & 171, 174 & Decay & $\Gamma_{rL}=1.02\times10^{4}\,\mathrm{s}^{-1}$\newline $51\%$ branch of a $50\,\mu\mathrm{s}$ Rydberg lifetime & \cite{Wilson2022,Senoo2025High} \\
    $\mathrm{DECAY}_{mg}$ & 171, 174 & Decay & $\Gamma_{mg}=(1\,\mathrm{s})^{-1}=1\,\mathrm{s}^{-1}$ & \cite{Lis2023,Siegel2024} \\
    $\mathrm{LOSS}_{g}$ & 171, 174 & Decay & $\Gamma_{gL}=(30\,\mathrm{s})^{-1}=0.033\,\mathrm{s}^{-1}$ & \cite{Jenkins2022,Muniz2025Repeated} \\
    $\mathrm{LOSS}_{m}$ & 171, 174 & Decay & $\Gamma_{mL}=(30\,\mathrm{s})^{-1}=0.033\,\mathrm{s}^{-1}$ & \cite{Lis2023} \\
    \midrule
    $\mathrm{LOSS}_{g}^{(\mathrm{hand})}$ & 171, 174 & Transportation & $p_{g\rightarrow L}^{(\mathrm{hand})}=0.1\%$ & \cite{Manetsch20246100,bluvstein2024logical,wang2026} \\
    $\mathrm{LOSS}_{m}^{(\mathrm{hand})}$ & 171, 174 & Transportation & $p_{m\rightarrow L}^{(\mathrm{hand})}=0.1\%$ & \cite{zhang2025lev} \\
    \bottomrule
\end{tabular}
\end{table*}

\subsection{Conventions}
\label{app:noise_conventions}

\quad All channels are specified by their Kraus error terms $K_{i>0}$, or by an equivalent measurement instrument when the channel includes a classical outcome. The ``no-error'' term $K_0$ satisfies the completeness relation
\begin{equation}
  K_0^{\dagger}K_0=\mathbb{I}-\sum_{i>0}K_i^\dagger K_i,
  \label{eq:kraus_no_error}
\end{equation}
where $\mathbb{I}$ is the identity over the whole modelled Hilbert space. We write $\Pi_a=\ket{a}\!\bra{a}$ and $\Pi_A=\sum_{a\in A}\ket{a}\!\bra{a}$ for the projector onto a state and a subspace, respectively. A transition $a\rightarrow b$ denotes the error term $\sqrt{p}\ket{b}\!\bra{a}$. More generally, $A\rightarrow B$ denotes the corresponding set of transitions between the two subspaces. For instance, for a single transition with probability~$p$,
\begin{equation}
  K_0=\mathbb{I}+\left(\sqrt{1-p}-1\right)\Pi_a,\qquad
  K_1=\sqrt{p}\ket{b}\!\bra{a}.
  \label{eq:kraus_transition}
\end{equation}

To avoid ambiguity in terminology, we distinguish the following terms according to the physical process being modelled. 
\begin{itemize}
  \item \emph{Loss} denotes one-way incoherent population transfer to the effective loss state $\ket{L}$, including trap loss associated with the finite trap lifetime and decay into dark states such as the $\tPt$ manifold. At the standard tweezer wavelength for Yb trapping, atoms that decay to the $\tPt$ state are rapidly lost from the trap~\cite{Ma2023high,Senoo2025High}. The state $\ket{L}$ represents all states outside the ground, metastable, and Rydberg manifolds retained in the model that are treated as unavailable for subsequent operations and readout. 

  \item \emph{Decay} denotes incoherent population transfer from excited states to lower-energy internal states, without distinguishing between mechanisms such as radiative decay and laser-induced photon scattering. It includes radiative decay from the Rydberg state to the metastable or ground manifold, as well as decay from the metastable manifold to the ground manifold. Here, loss is also classified as decay, although the specific term `loss' is preferred where applicable.

  \item The term \emph{leakage} refers to any process by which population is transferred from the computational subspace to non-computational states, that could later return to the computational subspace. Therefore, any form of loss cannot be a source of leakage by definition. This term is used for this phenomenon descriptively, and we refrain from using it to classify errors.

  \item Although the term \emph{damping} could be used where we use \emph{decay}, often the term `amplitude damping' is used to specifically refer to damping within the computational subspace. However, this is encoding dependent, since metastable-to-ground decay can be regarded as amplitude damping for the $\boson$ clock qubit but as erasure for the $\fermi$-m qubit. Hence, we opt to use the term decay, which is unambiguously independent of qubit encoding. We still use \emph{damping rates} to refer to coefficients of the Lindbladian, as is conventional.

\end{itemize}

Since we do not use information about atom loss or perform erasure detection, we do not use the term \textit{erasure} to refer to error sources throughout this text, except for describing components of the twirling approximation and when referring to related literature. We prefer using this term when erasure information is explicitly available, such as the location or timeframe of the erasure.

To specify the Pauli-error channels unambiguously, we first define the computational basis associated with each qubit encoding. For $\boson$, the optical clock-state basis is $\{\ket{g},\ket{m}\}\equiv\{\ket{0},\ket{1}\}$. For $\fermi$, the ground- and metastable-state bases are $\{\ket{g_0},\ket{g_1}\}$ and $\{\ket{m_0},\ket{m_1}\}$, respectively. We define $\Pi_c=\Pi_g+\Pi_m$, $\Pi_G=\Pi_{g_0}+\Pi_{g_1}$, and $\Pi_M=\Pi_{m_0}+\Pi_{m_1}$. Their complementary projectors are $\bar{\Pi}_c=\mathbb{I}-\Pi_c$, $\bar{\Pi}_G=\mathbb{I}-\Pi_G$, and $\bar{\Pi}_M=\mathbb{I}-\Pi_M$. Pauli operators act conventionally within the corresponding computational subspace and as the identity on its complement:
\begin{equation}
\begin{aligned}
  X_c&=\ket{g}\!\bra{m}+\ket{m}\!\bra{g}+\bar{\Pi}_c,\\
  Y_c&=-i\ket{g}\!\bra{m}+i\ket{m}\!\bra{g}+\bar{\Pi}_c,\\
  Z_c&=\Pi_g-\Pi_m+\bar{\Pi}_c,\\[2pt]
  X_g&=\ket{g_0}\!\bra{g_1}+\ket{g_1}\!\bra{g_0}+\bar{\Pi}_G,\\
  Y_g&=-i\ket{g_0}\!\bra{g_1}+i\ket{g_1}\!\bra{g_0}+\bar{\Pi}_G,\\
  Z_g&=\Pi_{g_0}-\Pi_{g_1}+\bar{\Pi}_G,\\[2pt]
  X_m&=\ket{m_0}\!\bra{m_1}+\ket{m_1}\!\bra{m_0}+\bar{\Pi}_M,\\
  Y_m&=-i\ket{m_0}\!\bra{m_1}+i\ket{m_1}\!\bra{m_0}+\bar{\Pi}_M,\\
  Z_m&=\Pi_{m_0}-\Pi_{m_1}+\bar{\Pi}_M.
\end{aligned}
\label{eq:pauli_bases}
\end{equation}
Two-qubit Pauli operators are tensor products of these full-space operators.

For clock-transition errors in $\fermi$, we also use pair-resolved Pauli operators acting on a selected ground--metastable pair $(\ket{g_j},\ket{m_k})$. Defining $\Pi_{g_jm_k}=\Pi_{g_j}+\Pi_{m_k}$ and $\bar{\Pi}_{g_jm_k}=\mathbb{I}-\Pi_{g_jm_k}$, these operators are
\begin{equation}
\begin{aligned}
 X_{g_j,m_k}
 &=\ket{g_j}\!\bra{m_k}+\ket{m_k}\!\bra{g_j}
   +\bar{\Pi}_{g_jm_k},\\
 Y_{g_j,m_k}
 &=-i\ket{g_j}\!\bra{m_k}+i\ket{m_k}\!\bra{g_j}
   +\bar{\Pi}_{g_jm_k},\\
 Z_{g_j,m_k}
 &=\Pi_{g_j}-\Pi_{m_k}+\bar{\Pi}_{g_jm_k}.
\end{aligned}
\label{eq:clock_transition_pauli}
\end{equation}
Thus, each operator has the usual Pauli action on the selected two-state subspace and acts as the identity on all other states. 

\subsection{Coherent-control channels}
\label{app:coherent_control_channels}

\quad The following channels describe errors associated with coherent control, including single-qubit rotations, clock-transition excitation, and Rydberg-mediated multi-qubit gates. They capture depolarising errors and incoherent population transfer arising from these driven processes.

\par\medskip\noindent{\bfseries\boldmath $\mathrm{DEP1}_{g}$ ($\fermi$).} This channel represents single-qubit depolarising error within the ground-state computational manifold of $\fermi$. The Kraus operators are $K_0=\sqrt{1-p_1^{(g)}}I$ and $K_P=\sqrt{p_1^{(g)}/3}\,P$ for $P\in\{X_g,Y_g,Z_g\}$.

\par\medskip\noindent{\bfseries\boldmath $\mathrm{DEP1}_{m}$ ($\fermi$).} This channel represents single-qubit depolarising error within the metastable-state computational manifold of $\fermi$. The Kraus operators are $K_0=\sqrt{1-p_1^{(m)}}I$ and $K_P=\sqrt{p_1^{(m)}/3}\,P$ for $P\in\{X_m,Y_m,Z_m\}$.

\par\medskip\noindent{\bfseries\boldmath $\mathrm{DEP1}_{gm}$ ($\fermi$).} This channel represents depolarising error during clock-transition excitation between the ground- and metastable-state manifolds of $\fermi$. Using the pair-resolved Pauli operators defined in Eq.~\eqref{eq:clock_transition_pauli}, the Kraus operators are
\begin{equation}
\begin{aligned}
  K_0&=\sqrt{1-p_1^{(gm)}}I,\\
  K_{P}^{(j,k)}
  &=\sqrt{\frac{p_1^{(gm)}}{12}}\,P_{g_j,m_k},
\end{aligned}
\end{equation}
where $P_{g_j,m_k}\in\{X_{g_j,m_k},Y_{g_j,m_k},Z_{g_j,m_k}\}$ and $j,k\in\{0,1\}$. The twelve error operators correspond to the three Pauli errors for each of the four ground--metastable state pairs.

\par\medskip\noindent{\bfseries\boldmath $\mathrm{DEP1}_{c}$ ($\boson$).} This channel represents single-qubit depolarising error on the $\sSz\leftrightarrow\tPz$ clock-transition of $\boson$. The Kraus operators are $K_0=\sqrt{1-p_1^{(c)}}I$ and $K_P=\sqrt{p_1^{(c)}/3}\,P$ for $P\in\{X_c,Y_c,Z_c\}$.

\par\medskip\noindent{\bfseries\boldmath $\mathrm{DECAY}_{mg}^{(\mathrm{gate})}$ ($\fermi$).} This channel describes metastable-to-ground-state decay induced by the $\tPz$ nuclear-spin control laser through off-resonant scattering via the $\tDo$ and $\tPo$ states~\cite{Senoo2025High}. As a modelling assumption, we neglect the lifetime of the intermediate state and represent the process as a direct transition with equal branching to the two ground-state qubit levels. The Kraus operators are
\begin{equation}
\begin{aligned}
  K_0&=I+\left(\sqrt{1-p_{m\rightarrow g}^{(\mathrm{gate})}}-1\right)\Pi_M,\\
  K^{(j,k)}&=\sqrt{\frac{p_{m\rightarrow g}^{(\mathrm{gate})}}{2}}
  \ket{g_j}\!\bra{m_k},
  \qquad j,k\in\{0,1\}.
\end{aligned}
\end{equation}

\par\medskip\noindent{\bfseries\boldmath $\mathrm{DEP2}_{m}$ ($\fermi$).} This channel represents two-qubit depolarising error during Rydberg-mediated gates between metastable-state qubits. Defining $\mathcal{P}_{2,m}^{\ast}=\{I,X_m,Y_m,Z_m\}^{\otimes2}\setminus\{I\otimes I\}$, the Kraus operators are $K_0=\sqrt{1-p_2^{(m)}}I\otimes I$ and $K_P=\sqrt{p_2^{(m)}/15}\,P$ for $P\in\mathcal{P}_{2,m}^{\ast}$.

\par\medskip\noindent{\bfseries\boldmath $\mathrm{DEP2}_{c}$ ($\boson$).} This channel represents two-qubit depolarising error during Rydberg-mediated gates between clock qubits. Defining $\mathcal{P}_{2,c}^{\ast}=\{I,X_c,Y_c,Z_c\}^{\otimes2}\setminus\{I\otimes I\}$, the Kraus operators are $K_0=\sqrt{1-p_2^{(c)}}I\otimes I$ and $K_P=\sqrt{p_2^{(c)}/15}\,P$ for $P\in\mathcal{P}_{2,c}^{\ast}$.

\par\medskip\noindent{\bfseries\boldmath $\mathrm{DEP2}_{\mathrm{dual}}$ ($\fermi$, $\boson$).} This channel represents two-qubit depolarising error during Rydberg-mediated gates between a $\fermi$-m qubit and a $\boson$ clock qubit. Defining
\begin{equation*}
  \mathcal{P}_{2,\mathrm{dual}}^{\ast}
  =
  \left(\{I,X_c,Y_c,Z_c\}\otimes
  \{I,X_m,Y_m,Z_m\}\right)
  \setminus\{I\otimes I\},
\end{equation*}
the Kraus operators are $K_0=\sqrt{1-p_2^{(\mathrm{dual})}}I\otimes I$ and $K_P=\sqrt{p_2^{(\mathrm{dual})}/15}\,P$ for $P\in\mathcal{P}_{2,\mathrm{dual}}^{\ast}$. In the simulations, we set $p_2^{(\mathrm{dual})}$ to be the same as the two-qubit depolarising-error probability used for the individual qubit types (see Table~\ref{tab:noise_channel_summary}). This target-performance assumption is motivated by recent progress in interspecies Rydberg gates and by the expectation that the dominant gate errors can be reduced with improved Rydberg-gate control~\cite{white2026, miles2026, wang2026stab}.

\subsection{Measurement channels}
\label{app:measurement_channels}

\quad The following channels describe errors associated with state detection. Quantum processes caused by the detection sequence, such as loss and population flips, are applied before the classical measurement-error model. We use state-selective readout in the simulation. The readout process consists of two fluorescence-imaging steps, with an additional clock de-excitation step for the $\fermi$ metastable-state and $\boson$ clock-state encodings (see Eqs.~\ref{eq:app:measurement_order_171g}, \ref{eq:app:measurement_order_171m}, and \ref{eq:app:measurement_order_174}). To isolate noise channels intrinsic to measurement, this subsection includes only noise sources associated with ground-state fluorescence imaging. Other noise sources, such as those arising during clock de-excitation, idling, and decay, are described separately in Appendices~\ref{app:coherent_control_channels}, \ref{app:idling_channels}, and \ref{app:decay_channels}, respectively.

\par\medskip\noindent{\bfseries\boldmath $\mathrm{LOSS}_{g}^{(\mathrm{meas})}$ ($\fermi$, $\boson$).} This channel represents atom loss from the ground-state manifold during fluorescence state detection. The Kraus operators for $\fermi$ are
\begin{equation}
\begin{aligned}
  K_0&=I+\left(\sqrt{1-p_{g\rightarrow L}^{(\mathrm{meas})}}-1\right)\Pi_G,\\
  K_{g_j}&=\sqrt{p_{g\rightarrow L}^{(\mathrm{meas})}}\ket{L}\!\bra{g_j},
  \qquad j\in\{0,1\},
\end{aligned}
\end{equation}
whereas those for $\boson$ are
\begin{equation}
\begin{aligned}
  K_0&=I+\left(\sqrt{1-p_{g\rightarrow L}^{(\mathrm{meas})}}-1\right)\Pi_g,\\
  K_1&=\sqrt{p_{g\rightarrow L}^{(\mathrm{meas})}}\ket{L}\!\bra{g}.
\end{aligned}
\end{equation}

\par\medskip\noindent{\bfseries\boldmath $\mathrm{FLIP}_{g}$ ($\fermi$).} This channel represents an in-manifold bit flip induced during measurement. Because the nuclear-spin states of the ground-state qubit are nearly degenerate, measurement lasers addressing the $\sSz\leftrightarrow\sPo$ and $\sSz\leftrightarrow\tPo$ transitions can induce bit-flip errors within this encoding. In contrast, the $\boson$ clock qubit is protected by the large energy separation between the $\sSz$ and $\tPz$ states. This channel is therefore applied only to $\fermi$.
The Kraus operators are $K_0=\sqrt{1-p_{\mathrm{flip}}^{(g)}}I$ and $K_1=\sqrt{p_{\mathrm{flip}}^{(g)}}X_g$.

\begin{table*}[!tbp]
  \centering
  \caption{Computational-state detection under noisy two-step readout. B and D denote bright and dark outcomes, respectively. For compactness, $p\equiv p_{\mathrm{meas}}$, where $p_{\mathrm{meas}}$ is the independent discrimination-error probability for each imaging step listed in Table~\ref{tab:noise_channel_summary}. 
  The entries are shown for the $\boson$ clock qubit. In the $\fermi$-g qubit case, $\ket{0}$ and $\ket{1}$ ordinarily correspond to $\ket{g_0}$ and $\ket{g_1}$.
  }
  \label{tab:readout_error}
  \small
  \renewcommand{\arraystretch}{1.12}
  \begin{tabular}{@{}p{0.10\textwidth}p{0.13\textwidth}p{0.10\textwidth}p{0.27\textwidth}p{0.22\textwidth}@{}}
    \toprule
    Actual state & Detected state & Ideal record &
      \shortstack[l]{Noisy record contributing\\to detection} &
      Probability \\
    \midrule
    $\ket{0}$ & $\ket{0}$ & BD
      & BD or BB detected as $\ket{0}$
      & $p_{0,0}=(1-p)^2+\frac{p(1-p)}{2}$ \\
    $\ket{0}$ & $\ket{1}$ & BD
      & DB or BB detected as $\ket{1}$
      & $p_{0,1}=p^2+\frac{p(1-p)}{2}$ \\
    $\ket{0}$ & $\ket{L}$ & BD
      & DD
      & $p_{0,L}=p(1-p)$ \\
    $\ket{1}$ & $\ket{0}$ & DB
      & BD or BB detected as $\ket{0}$
      & $p_{1,0}=p^2+\frac{p(1-p)}{2}$ \\
    $\ket{1}$ & $\ket{1}$ & DB
      & DB or BB detected as $\ket{1}$
      & $p_{1,1}=(1-p)^2+\frac{p(1-p)}{2}$ \\
    $\ket{1}$ & $\ket{L}$ & DB
      & DD
      & $p_{1,L}=p(1-p)$ \\
    $\ket{L}$ & $\ket{0}$ & DD
      & BD or BB detected as $\ket{0}$
      & $p_{L,0}=p(1-p)+\frac{p^2}{2}$ \\
    $\ket{L}$ & $\ket{1}$ & DD
      & DB or BB detected as $\ket{1}$
      & $p_{L,1}=p(1-p)+\frac{p^2}{2}$ \\
    $\ket{L}$ & $\ket{L}$ & DD
      & DD
      & $p_{L,L}=(1-p)^2$ \\
    \bottomrule
  \end{tabular}
\end{table*}

\par\medskip\noindent{\bfseries\boldmath $\mathrm{MERR}$ ($\fermi$, $\boson$).} This channel represents classical discrimination errors in loss-aware state-selective readout~\cite{Lis2023,Norcia2023,Huie2023,Li2025Fast,Senoo2025High,Baranes2026}.
The fluorescence imaging yields either a bright (B) or dark (D) outcome. The two-step records BD and DB are detected respectively as $\ket{g}$ and $\ket{m}$ for the $\boson$ clock qubit, and as $\ket{g_0}$ and $\ket{g_1}$ for the $\fermi$-g nuclear-spin qubit. The DD outcome is detected as loss $\ket{L}$. 
Since the BB outcome cannot be interpreted in a physical sense, when this outcome is detected it is treated as either computational-basis state with equal probability. Table~\ref{tab:readout_error} summarises the resulting detection events.

Let $p_{a,b}$ denote the probability listed in Table~\ref{tab:readout_error}
that the actual state $\ket{a}$ is detected as $\ket{b}$, where
$a,b\in\{g,m,L\}$ for the $\boson$ clock qubit and $a,b\in\{g_0,g_1,L\}$ for the $\fermi$-g nuclear-spin qubit. The corresponding Kraus operators are
\begin{equation}
\begin{aligned}
  &\qquad K^{(a,b)}=\sqrt{p_{a,b}}\,\ket{b}\!\bra{a}, \\
  a,b\in&
  \begin{cases}
      \{g,m,L\} ~\text{for the $\boson$ qubit}, \\
      \{g_0,g_1,L\} ~\text{for the $\fermi$-g qubit},
  \end{cases}
\end{aligned}
\end{equation}
with $\sum_{b}p_{a,b}=1$ for each actual state $a$.

\subsection{Reset channels}
\label{app:reset_channels}

\quad The following channels describe errors arising from the optical transitions used for internal-state preparation and motional reset. Ground-manifold reset and cooling involve photon scattering on transitions such as $\sSz\leftrightarrow\tPo$, whereas metastable-manifold reset involves excitation from $\tPz$ to an upper energy manifold followed by spontaneous decay. Repeated scattering can change the nuclear-spin state, while decay through untrapped or dark branches can transfer population out of the encoded manifold. We model these effects as in-manifold flips and reset-induced loss. Flip channels for $\boson$ are omitted because this isotope has no nuclear-spin degree of freedom and its clock states are separated by an optical-frequency energy gap.

\par\medskip\noindent{\bfseries\boldmath $\mathrm{LOSS}_{g}^{(\mathrm{reset})}$ ($\fermi$, $\boson$).} This channel represents loss from the ground-state manifold during either the $Z$-basis state preparation or motional reset. For $\fermi$, the Kraus operators are
\begin{equation}
\begin{aligned}
  K_0&=I+\left(\sqrt{1-p_{g\rightarrow L}^{(\mathrm{reset})}}-1\right)\Pi_G,\\
  K_{g_j}&=\sqrt{p_{g\rightarrow L}^{(\mathrm{reset})}}
  \ket{L}\!\bra{g_j},
  \qquad j\in\{0,1\}.
\end{aligned}
\end{equation}
For $\boson$, the Kraus operators are
\begin{equation}
\begin{aligned}
  K_0&=I+\left(\sqrt{1-p_{g\rightarrow L}^{(\mathrm{reset})}}-1\right)\Pi_g,\\
  K_1&=\sqrt{p_{g\rightarrow L}^{(\mathrm{reset})}}
  \ket{L}\!\bra{g}.
\end{aligned}
\end{equation}

\par\medskip\noindent{\bfseries\boldmath $\mathrm{LOSS}_{m}^{(\mathrm{reset})}$ ($\fermi$).} This channel represents loss from the metastable-state manifold via transitions to untrapped states or dark states accessed during spin reset~\cite{Ma2023high,zhang2025lev}. The Kraus operators are
\begin{equation}
\begin{aligned}
  K_0&=I+\left(\sqrt{1-p_{m\rightarrow L}^{(\mathrm{reset})}}-1\right)\Pi_M,\\
  K_{m_j}&=\sqrt{p_{m\rightarrow L}^{(\mathrm{reset})}}
  \ket{L}\!\bra{m_j},
  \qquad j\in\{0,1\}.
\end{aligned}
\end{equation}

\par\medskip\noindent{\bfseries\boldmath $\mathrm{FLIP}_{g}$ ($\fermi$).} This channel represents a nuclear-spin flip within the ground-state manifold caused by photon scattering during either the $Z$-basis state preparation or motional reset. The Kraus operators are $K_0=\sqrt{1-p_{\mathrm{flip}}^{(g)}}I$ and $K_1=\sqrt{p_{\mathrm{flip}}^{(g)}}X_g$.

\par\medskip\noindent{\bfseries\boldmath $\mathrm{FLIP}_{m}$ ($\fermi$).} This channel represents a nuclear-spin flip within the metastable-state manifold caused by photon scattering during the $Z$-basis state preparation. The Kraus operators are $K_0=\sqrt{1-p_{\mathrm{flip}}^{(m)}}I$ and $K_1=\sqrt{p_{\mathrm{flip}}^{(m)}}X_m$.

\subsection{Idling channels}
\label{app:idling_channels}

\quad The following channels describe dephasing and bit-flip errors that accumulate while a qubit is idle or being shuttled and is not subject to active control. They arise from finite coherence and population-relaxation times within the ground- and metastable-state manifolds, as well as finite clock-transition coherence between these manifolds. Unlike the operation-specific channels above, their error probabilities depend on the idling duration $t$ through $p(t)=1-e^{-t/T}$, where $T$ is the corresponding $T_1$ or $T_2$ timescale.
These channels are disabled during coherent-control operations and reset because the associated coherence and bit-flip errors are already incorporated into the operation-specific error channels; applying the idling channels simultaneously would therefore double count these contributions.

\par\medskip\noindent{\bfseries\boldmath $\mathrm{ZERR}_{g}$ ($\fermi$).} This channel represents a phase-flip error within the ground-state qubit. Defining $p_Z^{(g)}=1-e^{-t/T_2^{(g)}}$, the Kraus operators are $K_0=\sqrt{1-p_Z^{(g)}}I$ and $K_1=\sqrt{p_Z^{(g)}}Z_g$.

\par\medskip\noindent{\bfseries\boldmath $\mathrm{XERR}_{g}$ ($\fermi$).} This channel represents a bit-flip error within the ground-state qubit. Defining $p_X^{(g)}=1-e^{-t/T_1^{(g)}}$, the Kraus operators are $K_0=\sqrt{1-p_X^{(g)}}I$ and $K_1=\sqrt{p_X^{(g)}}X_g$.

\par\medskip\noindent{\bfseries\boldmath $\mathrm{ZERR}_{m}$ ($\fermi$).} This channel represents a phase-flip error within the metastable-state qubit. Defining $p_Z^{(m)}=1-e^{-t/T_2^{(m)}}$, the Kraus operators are $K_0=\sqrt{1-p_Z^{(m)}}I$ and $K_1=\sqrt{p_Z^{(m)}}Z_m$.

\par\medskip\noindent{\bfseries\boldmath $\mathrm{XERR}_{m}$ ($\fermi$).} This channel represents a bit-flip error within the metastable-state qubit. Defining $p_X^{(m)}=1-e^{-t/T_1^{(m)}}$, the Kraus operators are $K_0=\sqrt{1-p_X^{(m)}}I$ and $K_1=\sqrt{p_X^{(m)}}X_m$.

\par\medskip\noindent{\bfseries\boldmath $\mathrm{ZERR}_{gm}$ ($\fermi$).} This channel represents a phase-flip error associated with clock-transition coherence between the ground- and metastable-state manifolds. Defining $\bar{\Pi}_{GM}=I-\Pi_G-\Pi_M$, the full-space phase operator is
\begin{equation}
  Z_{gm}=\Pi_G-\Pi_M+\bar{\Pi}_{GM}.
\end{equation}
For an error probability $p_Z^{(gm)}=1-e^{-t/T_2^{(gm)}}$, the Kraus operators are
$K_0=\sqrt{1-p_Z^{(gm)}}I$ and
$K_1=\sqrt{p_Z^{(gm)}}Z_{gm}$.

\par\medskip\noindent{\bfseries\boldmath $\mathrm{ZERR}_{c}$ ($\boson$).} This channel represents a phase-flip error within the clock qubit. Defining $p_Z^{(c)}=1-e^{-t/T_2^{(c)}}$, the Kraus operators are $K_0=\sqrt{1-p_Z^{(c)}}I$ and $K_1=\sqrt{p_Z^{(c)}}Z_c$.

\subsection{Decay channels}
\label{app:decay_channels}

\quad The following channels describe decay and atom loss that can occur independently of the circuit operation. They arise from finite Rydberg- and metastable-state lifetimes, branching into other atomic manifolds or dark states, and finite trap lifetimes. Their error probabilities depend on the elapsed time $t$ according to $p(t)=1-e^{-\Gamma t}$, where $\Gamma$ is the corresponding decay or loss rate. For Rydberg-state decay, we use a total radiative decay rate $\Gamma_{\mathrm{Ryd}}$ and branch it into the ground-state, metastable-state ($\tPz$), and other manifolds with branching ratios 0.42, 0.07, and 0.51, respectively~\cite{Senoo2025High}. The effective branch-specific rates are therefore $\Gamma_{rg}=0.42\,\Gamma_{\mathrm{Ryd}}$, $\Gamma_{rm}=0.07\,\Gamma_{\mathrm{Ryd}}$, and $\Gamma_{rL}=0.51\,\Gamma_{\mathrm{Ryd}}$. In contrast to the operation-specific and idling channels, these channels remain active during all circuit operations.

\par\medskip\noindent{\bfseries\boldmath $\mathrm{DECAY}_{rg}$ ($\fermi$, $\boson$).} This channel represents radiative decay from the Rydberg state to the ground-state manifold. For $\fermi$, defining $p_{r\rightarrow g}=1-e^{-\Gamma_{rg}t}$ and assuming equal branching to the two ground-state qubit levels, the Kraus operators are
\begin{equation}
  \begin{aligned}
    K_0&=I+\left(\sqrt{1-p_{r\rightarrow g}}-1\right)\Pi_r,\\
    K_{g_s}&=\sqrt{\frac{p_{r\rightarrow g}}{2}}\ket{g_s}\!\bra{r},
    \quad s\in\{0,1\}.
  \end{aligned}
\end{equation}
For $\boson$, using the same $p_{r\rightarrow g}$, the Kraus operators are
\begin{equation}
  \begin{aligned}
    K_0&=I+\left(\sqrt{1-p_{r\rightarrow g}}-1\right)\Pi_r,\\
    K_1&=\sqrt{p_{r\rightarrow g}}\ket{g}\!\bra{r}.
  \end{aligned}
\end{equation}

\par\medskip\noindent{\bfseries\boldmath $\mathrm{DECAY}_{rm}$ ($\fermi$, $\boson$).} This channel represents spontaneous decay from the Rydberg state to the metastable-state manifold. For $\fermi$, defining $p_{r\rightarrow m}=1-e^{-\Gamma_{rm}t}$ and assuming equal branching to the two metastable-state qubit levels, the Kraus operators are
\begin{equation}
  \begin{aligned}
    K_0&=I+\left(\sqrt{1-p_{r\rightarrow m}}-1\right)\Pi_r,\\
    K_{m_s}&=\sqrt{\frac{p_{r\rightarrow m}}{2}}\ket{m_s}\!\bra{r},
    \quad s\in\{0,1\}.
  \end{aligned}
\end{equation}
For $\boson$, using the same $p_{r\rightarrow m}$, the Kraus operators are
\begin{equation}
  \begin{aligned}
    K_0&=I+\left(\sqrt{1-p_{r\rightarrow m}}-1\right)\Pi_r,\\
    K_1&=\sqrt{p_{r\rightarrow m}}\ket{m}\!\bra{r}.
  \end{aligned}
\end{equation}

\par\medskip\noindent{\bfseries\boldmath $\mathrm{LOSS}_{r}$ ($\fermi$, $\boson$).} This channel represents Rydberg-state decay into untrapped or dark states. For both $\fermi$ and $\boson$, defining $p_{r\rightarrow L}=1-e^{-\Gamma_{rL}t}$, the Kraus operators are
\begin{equation}
  \begin{aligned}
    K_0&=I+\left(\sqrt{1-p_{r\rightarrow L}}-1\right)\Pi_r,\\
    K_1&=\sqrt{p_{r\rightarrow L}}\ket{L}\!\bra{r}.
  \end{aligned}
\end{equation}

\par\medskip\noindent{\bfseries\boldmath $\mathrm{DECAY}_{mg}$ ($\fermi$, $\boson$).} This channel represents decay from the metastable-state manifold to the ground-state manifold. Although the natural metastable-state lifetime exceeds $20\,\mathrm{s}$~\cite{Porsev2004}, off-resonant scattering from a 759~nm tweezer beam can transfer population from the $\tPz$ state to the $\sSz$ ground state through the $\tSo$ and $\tPo$ states, thereby shortening the effective metastable-state lifetime~\cite{Lis2023,Siegel2024}. This effect is especially dominant during imaging, where a deep trap is required. Since this scattering-induced decay rate is proportional to the trap depth~\cite{Siegel2024}, its value can differ between imaging and computation. For simplicity, we use a constant decay rate throughout the simulations, as listed in Table~\ref{tab:noise_channel_summary}. 
The resulting metastable-to-ground-state decay is modelled as a direct transition with equal branching to the two ground-state qubit levels.
For $\fermi$, defining $p_{m\rightarrow g}=1-e^{-\Gamma_{mg}t}$  and assuming equal branching to the two ground-state qubit levels, the Kraus operators are
\begin{equation}
  \begin{aligned}
    K_0&=I+\left(\sqrt{1-p_{m\rightarrow g}}-1\right)\Pi_M,\\
    K^{(j,k)}
      &=\sqrt{\frac{p_{m\rightarrow g}}{2}}\ket{g_j}\!\bra{m_k},
      \quad j,k\in\{0,1\}.
  \end{aligned}
\end{equation}
For $\boson$, using the same $p_{m\rightarrow g}$, the Kraus operators are
\begin{equation}
  \begin{aligned}
    K_0&=I+\left(\sqrt{1-p_{m\rightarrow g}}-1\right)\Pi_m,\\
    K_1&=\sqrt{p_{m\rightarrow g}}\ket{g}\!\bra{m}.
  \end{aligned}
\end{equation}

\par\medskip\noindent{\bfseries\boldmath $\mathrm{LOSS}_{g}$ ($\fermi$, $\boson$).} This channel represents loss from the ground-state manifold to $\ket{L}$ due to the finite trap lifetime. The trap lifetime depends on the background-gas collision rate~\cite{Schymik2021,Pichard2024Cryo,Zhang2025} and the trap depth. For $\fermi$, defining $p_{g\rightarrow L}=1-e^{-\Gamma_{gL}t}$, the Kraus operators are
\begin{equation}
  \begin{aligned}
    K_0&=I+\left(\sqrt{1-p_{g\rightarrow L}}-1\right)\Pi_G,\\
    K_{g_s}
      &=\sqrt{p_{g\rightarrow L}}\ket{L}\!\bra{g_s},
      \quad s\in\{0,1\}.
  \end{aligned}
\end{equation}
For $\boson$, using the same $p_{g\rightarrow L}$, the Kraus operators are
\begin{equation}
  \begin{aligned}
    K_0&=I+\left(\sqrt{1-p_{g\rightarrow L}}-1\right)\Pi_g,\\
    K_1&=\sqrt{p_{g\rightarrow L}}\ket{L}\!\bra{g}.
  \end{aligned}
\end{equation}

\par\medskip\noindent{\bfseries\boldmath $\mathrm{LOSS}_{m}$ ($\fermi$, $\boson$).} This channel represents loss from the metastable-state manifold to $\ket{L}$ due to the finite trap lifetime. As in the ground-state manifold, the trap lifetime depends on the background-gas collision rate and the trap depth. In a magic-wavelength trap, where the ground and metastable states experience the same optical trapping potential~\cite{Katori2003, Takamoto2003}, 
we take the metastable-state trap lifetime to be the same as the ground-state trap lifetime. For $\fermi$, defining $p_{m\rightarrow L}=1-e^{-\Gamma_{mL}t}$, the Kraus operators are
\begin{equation}
  \begin{aligned}
    K_0&=I+\left(\sqrt{1-p_{m\rightarrow L}}-1\right)\Pi_M,\\
    K_{m_s}
      &=\sqrt{p_{m\rightarrow L}}\ket{L}\!\bra{m_s},
      \quad s\in\{0,1\}.
  \end{aligned}
\end{equation}
For $\boson$, using the same $p_{m\rightarrow L}$, the Kraus operators are
\begin{equation}
  \begin{aligned}
    K_0&=I+\left(\sqrt{1-p_{m\rightarrow L}}-1\right)\Pi_m,\\
    K_1&=\sqrt{p_{m\rightarrow L}}\ket{L}\!\bra{m}.
  \end{aligned}
\end{equation}
The rate $\Gamma_{mL}$ used here is estimated from the finite trap lifetime and does not include additional $\tPz$ loss induced by tweezer-light photon scattering. Such scattering can transfer the $\tPz$ state into dark manifolds such as $\tPt$~\cite{Lis2023,Siegel2024}, thereby increasing the effective loss rate. This effect is not included in the present simulations and would primarily affect configurations that rely on metastable-state shelving.

\subsection{Transportation channels}
\label{app:transportation_channels}

\quad The following channels describe errors associated with atom transportation. The transport process can introduce loss when atoms are handed over between a static trapping array and the dynamically movable trap used for shuttling. To isolate noise channels intrinsic to the transportation process,
this subsection includes only noise sources associated with handover loss. Time-dependent errors accumulated during transport, including dephasing, spin relaxation, metastable-state decay, and finite trap lifetime, are separately characterised by the idling and decay channels using the transport time (see Appendices ~\ref{app:idling_channels} and \ref{app:decay_channels}, respectively). We model the handover error as a loss channel from the ground- or metastable-state manifold to $\ket{L}$. For simplicity, we ignore heating during transportation~\cite{saffman2025} and decoherence during handover~\cite{zhang2025lev}.

\par\medskip\noindent{\bfseries\boldmath $\mathrm{LOSS}_{g}^{(\mathrm{hand})}$ ($\fermi$, $\boson$).} This channel represents atom loss from the ground-state manifold during handover between static and dynamically movable traps. For $\fermi$, the Kraus operators are
\begin{equation}
  \begin{aligned}
    K_0&=I+\left(\sqrt{1-p_{g\rightarrow L}^{(\mathrm{hand})}}-1\right)\Pi_G,\\
    K_{g_s}
      &=\sqrt{p_{g\rightarrow L}^{(\mathrm{hand})}}\ket{L}\!\bra{g_s},
      \quad s\in\{0,1\}.
  \end{aligned}
\end{equation}
For $\boson$, the Kraus operators are
\begin{equation}
  \begin{aligned}
    K_0&=I+\left(\sqrt{1-p_{g\rightarrow L}^{(\mathrm{hand})}}-1\right)\Pi_g,\\
    K_1&=\sqrt{p_{g\rightarrow L}^{(\mathrm{hand})}}\ket{L}\!\bra{g}.
  \end{aligned}
\end{equation}

\par\medskip\noindent{\bfseries\boldmath $\mathrm{LOSS}_{m}^{(\mathrm{hand})}$ ($\fermi$, $\boson$).} This channel represents atom loss from the metastable-state manifold during handover between static and dynamically movable traps. For $\fermi$, the Kraus operators are
\begin{equation}
  \begin{aligned}
    K_0&=I+\left(\sqrt{1-p_{m\rightarrow L}^{(\mathrm{hand})}}-1\right)\Pi_M,\\
    K_{m_s}
      &=\sqrt{p_{m\rightarrow L}^{(\mathrm{hand})}}\ket{L}\!\bra{m_s},
      \quad s\in\{0,1\}.
  \end{aligned}
\end{equation}
For $\boson$, the Kraus operators are
\begin{equation}
  \begin{aligned}
    K_0&=I+\left(\sqrt{1-p_{m\rightarrow L}^{(\mathrm{hand})}}-1\right)\Pi_m,\\
    K_1&=\sqrt{p_{m\rightarrow L}^{(\mathrm{hand})}}\ket{L}\!\bra{m}.
  \end{aligned}
\end{equation}

\subsection{Gate time and idling time}
\label{app:gate_idling_time}

\quad The numerical model accounts for operation times from reset in a memory experiment until state readout is complete. The operation times determine the exposure time for idling, decay, and transportation channels. Table~\ref{tab:gate_idling_times} summarises the values used in the simulations.

Transport-schedule design aimed at suppressing motional heating during atom transport has been widely investigated~\cite{bluvstein2022,Finkelstein2024,Manetsch20246100,Pagano2024,Hwang2025,Chinnarasu2025,Cicali2025,zhang2025lev}. In this simulation, we use a constant-acceleration trajectory to simplify the estimate of transport times. For a shuttling distance $l$, the transport time is
\begin{equation}
  t_{\mathrm{move}}(l)=2\sqrt{\frac{l}{a}},
\end{equation}
where $a$ is the acceleration in the first half of the transport and the deceleration in the second half. Since transport is used only for zoned readout in this simulation, the transport distance is set to $l = d \times l_{\mathrm{site}} + l_{\mathrm{zone}}$, where $d$ is the code distance of the surface code, and $l_{\mathrm{site}}$ and $l_{\mathrm{zone}}$ are the site spacing and zoned separation, respectively. The usable transport time is limited by acoustic-lensing effects at large velocities~\cite{Manetsch20246100,zhang2025lev}, which can distort the moving trap and increase trap loss. In addition, handover between static and movable traps is included in the simulation. The handover time is limited by the adiabatic trap-depth ramp used to suppress heating during transfer.

\begin{table}[htbp]
  \centering
  \small
  \setlength{\tabcolsep}{3pt}
  \renewcommand{\arraystretch}{1.08}
  \caption{Operation times and transport parameters used in the numerical model. The values determine the time intervals over which the corresponding time-dependent noise channels are accumulated.}
  \label{tab:gate_idling_times}
  \begin{tabular}{@{}llll@{}}
    \toprule
    \tablelcell{0.30\columnwidth}{Operation or parameter}
      & \tableccell{0.17\columnwidth}{Notation}
      & \tableccell{0.22\columnwidth}{Value used}
      & \tablelcell{0.23\columnwidth}{Refs.} \\
    \midrule
    \tablelcell{0.30\columnwidth}{Single-qubit gate\\($\boson$)}
      & \tableccell{0.17\columnwidth}{$t_{1Q}^{(c)}$}
      & \tableccell{0.22\columnwidth}{$100\,\mathrm{\mu s}$}
      & \tablelcell{0.23\columnwidth}{\cite{Finkelstein2024}} \\
    \tablelcell{0.30\columnwidth}{Single-qubit gate\\($\fermi$ clock)}
      & \tableccell{0.17\columnwidth}{$t_{1Q}^{(gm)}$}
      & \tableccell{0.22\columnwidth}{$10\,\mathrm{\mu s}$}
      & \tablelcell{0.23\columnwidth}{\cite{Lis2023}} \\
    \tablelcell{0.30\columnwidth}{Single-qubit gate\\g- or m-qubit\\($\fermi$)}
      & \tableccell{0.17\columnwidth}{$t_{1Q}^{(g)}, t_{1Q}^{(m)}$}
      & \tableccell{0.22\columnwidth}{$1\,\mathrm{\mu s}$}
      & \tablelcell{0.23\columnwidth}{\cite{Jenkins2022,Ma2022,Lis2023,AC2025High,Ma2023high,Li2025Parallelized}} \\
    \tablelcell{0.30\columnwidth}{Two-qubit gate}
      & \tableccell{0.17\columnwidth}{$t_{2Q}^{(c)}, t_{2Q}^{(m)}$\\$t_{2Q}^{(dual)}$}
      & \tableccell{0.22\columnwidth}{$0.3\,\mathrm{\mu s}$}
      & \tablelcell{0.23\columnwidth}{\cite{Senoo2025High,AC2025High,liu2026high}} \\
    \midrule
    \tablelcell{0.30\columnwidth}{Readout}
      & \tableccell{0.17\columnwidth}{$t_{\mathrm{read}}$}
      & \tableccell{0.22\columnwidth}{$1\,\mathrm{ms}$}
      & \tablelcell{0.23\columnwidth}{\cite{zhang2025lev,Senoo2025High,Li2025Parallelized,AC2026toric,Falconi2025,yokoyama2026}} \\
    \midrule
    \tablelcell{0.30\columnwidth}{Reset}
      & \tableccell{0.17\columnwidth}{$t_{\mathrm{reset}}$}
      & \tableccell{0.22\columnwidth}{$2\,\mathrm{ms}$}
      & \tablelcell{0.23\columnwidth}{\cite{Lis2023,Muniz2025Repeated,Li2025Parallelized,AC2026toric}} \\
    \midrule
    \tablelcell{0.30\columnwidth}{Transport\\acceleration}
      & \tableccell{0.17\columnwidth}{$a$}
      & \tableccell{0.22\columnwidth}{$5500\,\mathrm{m\,s^{-2}}$}
      & \tablelcell{0.23\columnwidth}{\cite{bluvstein2022, Zhou2025Resource}} \\
    \tablelcell{0.30\columnwidth}{Site spacing}
      & \tableccell{0.17\columnwidth}{$l_{\mathrm{site}}$}
      & \tableccell{0.22\columnwidth}{$3\,\mathrm{\mu m}$}
      & \tablelcell{0.23\columnwidth}{\cite{Norcia2024Ite}} \\
    \tablelcell{0.30\columnwidth}{Zoned separation}
      & \tableccell{0.17\columnwidth}{$l_{\mathrm{zone}}$}
      & \tableccell{0.22\columnwidth}{$100\,\mathrm{\mu m}$}
      & \tablelcell{0.23\columnwidth}{\cite{bluvstein2024logical,Bluvstein2025-ge}} \\
    \tablelcell{0.30\columnwidth}{Handover}
      & \tableccell{0.17\columnwidth}{$t_{\mathrm{hand}}$}
      & \tableccell{0.22\columnwidth}{$200\,\mathrm{\mu s}$}
      & \tablelcell{0.23\columnwidth}{\cite{bluvstein2024logical,Manetsch20246100,wang2026}} \\
    \bottomrule
  \end{tabular}
\end{table}

\subsection{Noise channel ordering definition}
\label{app:channel_ordering}

\quad This subsection defines the order in which active channels are composed within each circuit operation. Unless stated otherwise, an ideal operation is applied first, followed by operation-specific channels, time-dependent channels, and then the decay-channel block. Inactive channels are omitted for encodings or operations to which they do not apply. We use $U$ for an ideal unitary operation, $\widetilde{U}$ for the corresponding noisy operation, and $\widetilde{\mathcal{D}}$ for the decay-channel block defined below. For completeness, we explicitly list the ordering for the operations associated with the noise channels in Table~\ref{tab:noise_channel_summary}.

\par\medskip\noindent\textbf{Coherent-control operations.} Coherent-control gate operations follow the common rule, with the coherent-control channels treated as the operation-specific block. As an example, a dual-Yb two-qubit gate between a $\fermi$-m qubit and a $\boson$ clock qubit is written as
\begin{equation}
  \widetilde{U}_{2Q}^{(\mathrm{dual})} = \widetilde{\mathcal{D}} \circ \mathrm{DEP2}_{\mathrm{dual}} \circ U_{2Q}^{(\mathrm{dual})},
\end{equation}
where channels act from right to left. In this two-qubit example, each applicable one-qubit channel inside $\widetilde{\mathcal{D}}$ acts on both atoms; for example, $\mathrm{LOSS}_{r}$ abbreviates $\mathrm{LOSS}_{r}^{(171)}\otimes\mathrm{LOSS}_{r}^{(174)}$. 

As a second example, a single-qubit gate on the $m$ qubit has a coherent-control block containing both unitary depolarising noise and decay. In this case, the metastable-to-ground-state decay is applied after the depolarising channel and before the decay-channel block:
\begin{equation}
  \widetilde{U}_{1Q}^{(m)} = \widetilde{\mathcal{D}} \circ \mathrm{DECAY}_{mg}^{(\mathrm{gate})} \circ \mathrm{DEP1}_{m} \circ U_{1Q}^{(m)} .
\end{equation}

\par\medskip\noindent\textbf{Measurement.} Measurement consists of multiple state-selective operations, and the ordering depends on the encoding. We denote an ideal clock-state de-excitation by $D_c$. The measurement channel $M$ and the discrimination error $\mathrm{MERR}$ are applied at the end of the sequence of channels. During this time, idling errors accumulate on unmeasured qubits. For the $\fermi$-g qubit, the noisy measurement sequence is
\begin{equation}
  \begin{aligned}
  \widetilde{M}^{(171,g)} ={}& M^{(171,g)} \circ  \mathrm{MERR}\\
  & \circ \bigl( \widetilde{\mathcal{D}}\circ \mathrm{LOSS}_{g}^{(\mathrm{meas})}\circ\mathrm{FLIP}_{g}  \bigr) \\
  &\circ \bigl( \widetilde{\mathcal{D}}\circ \mathrm{LOSS}_{g}^{(\mathrm{meas})}\circ\mathrm{FLIP}_{g} \bigr),
  \end{aligned}
  \label{eq:app:measurement_order_171g}
\end{equation}
where the two factors in parentheses correspond to the two fluorescence imaging steps that detect population in the $\ket{g_0}$ and $\ket{g_1}$ states, respectively. 
For the $\fermi$-m qubit, the noisy measurement sequence is
\begin{equation}
  \begin{aligned}
  \widetilde{M}^{(171,m)} ={}& M^{(171,g)} \circ \mathrm{MERR}\\
  & \circ \bigl( \widetilde{\mathcal{D}}\circ \mathrm{LOSS}_{g}^{(\mathrm{meas})}\circ\mathrm{FLIP}_{g} \bigr)\\
  &\circ \bigl( \widetilde{\mathcal{D}}\circ\mathrm{DEP1}_{gm}\circ D_c \bigr) \\
  &\circ \bigl( \widetilde{\mathcal{D}}\circ \mathrm{LOSS}_{g}^{(\mathrm{meas})}\circ\mathrm{FLIP}_{g} \bigr) \\
  &\circ \bigl( \widetilde{\mathcal{D}}\circ \mathrm{DEP1}_{gm}\circ D_c \bigr),
  \end{aligned}
  \label{eq:app:measurement_order_171m}
\end{equation}
where the two $D_c$ factors describe clock-state de-excitation steps, and the two measurement factors describe fluorescence imaging of the $\ket{g_0}$ and $\ket{g_1}$ states after de-excitation. 
For the $\boson$ clock qubit, the noisy measurement sequence is
\begin{equation}
  \begin{aligned}
  \widetilde{M}^{(174)} ={}& M^{(174)} \circ \mathrm{MERR} \\
  & \circ \bigl( \widetilde{\mathcal{D}}\circ \mathrm{LOSS}_{g}^{(\mathrm{meas})} \bigr) \\
  &\circ \bigl( \widetilde{\mathcal{D}}\circ\mathrm{DEP1}_{c}\circ D_c \bigr) \\
  &\circ \bigl( \widetilde{\mathcal{D}}\circ \mathrm{LOSS}_{g}^{(\mathrm{meas})} \bigr), 
  \end{aligned}
  \label{eq:app:measurement_order_174}
\end{equation}
where the two measurement factors describe fluorescence imaging of the ground-state population before and after exchanging the $\ket{g}$ and $\ket{m}$ populations. 
Note that noise associated with the shelving of unmeasured qubits is not included in the above sequences. Depending on the qubit encoding, it is added before this sequence.

\par\medskip\noindent\textbf{Reset.} Reset operations depend on the encoding. We denote the ideal reset operations by $R_g^{(171)}$, $R_m^{(171)}$, and $R^{(174)}$ for the $\fermi$-g qubit, $\fermi$-m qubit and $\boson$ clock qubit, respectively. For the $\fermi$-g qubit, the noisy reset operation includes both nuclear-spin flips and ground-state reset loss:
\begin{equation}
  \widetilde{R}_{g}^{(171)} = \widetilde{\mathcal{D}} \circ \mathrm{LOSS}_{g}^{(\mathrm{reset})} \circ \mathrm{FLIP}_{g} \circ R_g^{(171)} .
\end{equation}
For the $\fermi$-m qubit, the noisy reset operation includes both nuclear-spin flips and metastable-state loss:
\begin{equation}
  \widetilde{R}_{m}^{(171)} = \widetilde{\mathcal{D}} \circ \mathrm{LOSS}_{m}^{(\mathrm{reset})} \circ \mathrm{FLIP}_{m} \circ R_m^{(171)} .
\end{equation}
For the $\boson$ clock qubit, nuclear-spin flips are absent, so the noisy reset operation contains only ground-state reset loss and the decay-channel block:
\begin{equation}
  \widetilde{R}^{(174)} = \widetilde{\mathcal{D}} \circ \mathrm{LOSS}_{g}^{(\mathrm{reset})} \circ R^{(174)} .
\end{equation}

\par\medskip\noindent\textbf{Idling.} For idling periods, $\mathrm{ZERR}$ is applied before $\mathrm{XERR}$, so the noisy idling operation is
\begin{equation}
  \widetilde{I} = \mathrm{XERR}\circ\mathrm{ZERR}\circ I .
\end{equation}

\par\medskip\noindent\textbf{Decay channels.} Whenever the decay-channel block is inserted, its internal ordering is fixed by decreasing damping rate:
\begin{equation}
  \begin{aligned}
  \widetilde{\mathcal{D}} ={}& \mathrm{LOSS}_{m} \circ \mathrm{LOSS}_{g} \circ \mathrm{DECAY}_{mg} \\
  &\circ \mathrm{DECAY}_{rm} \circ \mathrm{DECAY}_{rg} \circ \mathrm{LOSS}_{r}.
  \end{aligned}
\end{equation}

\par\medskip\noindent\textbf{Transportation.} Transport changes the atom position but not the encoded internal state, so the ideal operation is treated as the identity $I^{(\mathrm{trans})}$. The noisy transport operation is composed as
\begin{equation}
  \widetilde{I}^{(\mathrm{trans})} = \mathrm{LOSS}_{\mathrm{out}}^{(\mathrm{hand})} \circ \widetilde{\mathcal{D}} \circ \widetilde{I} \circ \mathrm{LOSS}_{\mathrm{in}}^{(\mathrm{hand})}  \circ I^{(\mathrm{trans})},
\end{equation}
where the two handover-loss channels correspond to transfer into and out of the movable trap, and errors involving noisy idling $\widetilde{I}$ and decay $\widetilde{\mathcal{D}}$ occur during shuttling.

\subsection{Justification of the decay channel ordering}
\label{app:channel_ordering_verification}

\quad Since decay channels act in a state-dependent manner, decomposing an aggregate decay process into individual channels cannot be arbitrary and requires a physically motivated ordering. For example, population initially in Rydberg states can reach the ground-state manifold through $\mathrm{DECAY}_{rm}$ followed by $\mathrm{DECAY}_{mg}$, whereas the reverse ordering does not describe the same physical pathway. 
This subsection focuses on the derivation and ordering of the composite decay channels that we have considered. Throughout the main text, decay has been treated as a separate process that occurs as a function of time, independent of ongoing operations. This assumption continues to hold throughout this subsection.

Choosing to formulate the decay process as a series of independent decay channels (rather than a single aggregate decay channel) is motivated by the error suppression budgets of Sec.~\ref{subsec:error_budget}. Each error term needs to be varied independently in order to evaluate each decay pathway's specific contribution to the budget. Nevertheless, as long as the composition of these independent channels gives the same aggregate decay channel, the specific choice of factorisation is mostly arbitrary.

First, let us derive the general decay channel via the Lindbladian. $H=0$ here by our assumptions. Letting the damping rates from higher energy to lower energy levels $i\to j$ be $\gamma_{ij}$ for each jump operator $|j\rangle\!\langle i|$, the following expression for the evolution of $\rho = \sum_{a,b}\rho_{ab}|a\rangle\!\langle b|$ can be derived:
\begin{align}
\dot{\rho}(t) = &\sum_j \Big(-\Gamma_j\rho_{jj} +\sum_{i>j}\gamma_{ij}\rho_{ii}\Big)|j\rangle\!\langle j|\nonumber\\ 
&-\frac{1}{2}\sum_{a\neq b} (\Gamma_a+\Gamma_b)\rho_{ab}|a\rangle\!\langle b|,
\label{eq:lindblad}
\end{align}
where $\Gamma_i$ is the decay rate of energy level $i$, that is, $\Gamma_i = \sum_{i>j}\gamma_{ij}$. We also write $\gamma_{ij} = b_j^{(i)}\Gamma_i$ where $b_j^{(i)}$ is the branching ratio for decay to energy level~$j$ from~$i$. 

Eq.~\eqref{eq:lindblad} shows that the off-diagonal terms of $\rho$ (between energy levels) can be considered independently. Therefore, $\rho_{ab}(t) = e^{-0.5(\Gamma_a+\Gamma_b)t}\rho_{ab}(0)$ for $a\neq b$. On the other hand, the dynamics of the diagonal terms can be understood in terms of transfer rates between the populations of each energy level $P_i = \rho_{ii}$. This can be solved with an appropriate linear equation:
\begin{equation}
\dot{P}(t) = AP \implies P(t) = e^{At}P(0), 
\end{equation}
where $A$ is the upper triangular matrix $A=\sum_j -\Gamma_j|j\rangle\!\langle j| +\sum_{i>j}\gamma_{ij}|j\rangle\!\langle i|$. Explicitly, for our case:
\begin{equation} 
A = \begin{bmatrix}
0 & \gamma_{gL} & \gamma_{mL}& \gamma_{rL}\\ 
0 & -\Gamma_g & \gamma_{mg}& \gamma_{rg}\\ 
0 & 0 & -\Gamma_m& \gamma_{rm}\\ 
0 & 0 & 0& -\Gamma_r
\end{bmatrix}.
\end{equation}

We want to map the dynamics described by the Lindbladian to a set of Kraus operators representing the aggregate decay channel i.e. a channel with $K_{i j} = \sqrt{p_{i\to j}} |j\rangle\!\langle i|$ and $K_0 = \sqrt{I - \sum_{i,j}K^\dag_{ij}K_{i j}}$. 
The jump operators $K_{ij}$ transfer population from $\ket{i}$ to $\ket{j}$, so $K_{ij}\rho K_{ij}^\dag = \rho_{ii}|j\rangle\!\langle j|$. Hence, each jump operator only contributes to the population terms found on the diagonal of the final state. Otherwise, the off-diagonal components of the final state are solely determined by the no-jump term: $K_0\rho K_0^\dag$.
Since $K_0$ is diagonal, we must take $K_0 = \sum_{a} \sqrt{e^{-\Gamma_at}}|a\rangle\!\langle a|$ in order to reproduce the off-diagonal terms $\rho_{ab}(t) = e^{-0.5(\Gamma_a+\Gamma_b)t}\rho_{ab}(0)$ from the Lindbladian. This choice for $K_0$ also exactly coincides with the diagonal terms of $e^{At}$.
The problem of solving for the remaining $K_{i\to j}$ then neatly reduces to the problem of solving for the off-diagonal terms of $e^{At}$, explicitly that is, $p_{i\to j}= \bra{j}e^{At}\ket{i}$.

To later exploit the fact that $\Gamma_r \gg \Gamma_m, \Gamma_g $, we shall separate the fast Rydberg decay from the slower ground- and metastable-state decay dynamics. Thus, we express $A$ as 
\begin{equation}
A = \begin{bmatrix}
L & \Gamma_r \vec{b}\\
0 & -\Gamma_r
\end{bmatrix},
\end{equation}
where $\vec{b}^T = (b_L^{(r)},b_g^{(r)},b_m^{(r)})$, the branching ratios for Rydberg decay. Solving gives the following expression:
\begin{equation}
e^{At} = \begin{bmatrix}
e^{Lt} & e^{Lt}(I-e^{-\Gamma_r(I+\frac{L}{\Gamma_r})t})(I+\frac{L}{\Gamma_r})^{-1}  \vec{b}\\
0 & e^{-\Gamma_rt}
\end{bmatrix}.
\end{equation}

From this we make the approximation that $(I+\frac{L}{\Gamma_r})\approx I$ since $\Gamma_r\gg|L|$, which gives:
\begin{align}e^{At} &\approx
\begin{bmatrix}
e^{Lt} & e^{Lt}(I-e^{-\Gamma_rt}) \vec{b}\\
0 & e^{-\Gamma_rt} 
\end{bmatrix}\\ &=
\begin{bmatrix}
e^{Lt} & 0\\
0 & 1 
\end{bmatrix}
\begin{bmatrix}
I & (I-e^{-\Gamma_rt}) \vec{b}\nonumber\\
0 & e^{-\Gamma_rt} 
\end{bmatrix},
\end{align}
where these two matrices are the solutions for if the Rydberg decay and lower energy levels had been decoupled and solved independently. Effectively, this means the aggregate channel is well-approximated by an independent fast Rydberg decay channel, dumping its population in the lower energy levels in ratio $\vec{b}$, followed by the slower decays encapsulated by~$L$. From here it is straightforward to choose an ordering for the factorised channels, dividing each half of the aggregate channel into three independent decays.

For the lower energy level decays:
\begin{equation}
e^{Lt} = \begin{bmatrix}
1 & 1-e^{-\gamma_{gL}t} & 1-e^{-\gamma_{mL}t}\\ 
0 & e^{-\Gamma_gt} & e^{-\gamma_{mL}t}(1-e^{-\gamma_{mg}t})\\ 
0 & 0 & e^{-\Gamma_mt}
\end{bmatrix}.
\end{equation}

This is achieved by three channels with $p_{m\to L} = 1-e^{-\gamma_{mL}t}$, $p_{g\to L} = 1-e^{-\gamma_{gL}t}$, and $p_{m\to  g} = 1-e^{-\gamma_{mg}t}$, respectively, applied in the order they have been stated. Note that because $\gamma_{gL} = \gamma_{mL}$ it can also be composed in the reversed order, which is what we choose here.

The aggregate Rydberg decay channel is described by $\hat{p}_{r \to i} = b^{(r)}_i(1-e^{-\Gamma_r t})$. The only consideration when factoring this channel into three constituents is that prior channels reduce the population acted on by later channels, reducing their effective rates. This is resolved by renormalising each probability by the probabilities of previous channels. We choose $p_{r\to L}= \hat{p}_{r\to L}$, $p_{r\to g}= \hat{p}_{r\to g}/(1-\hat{p}_{r\to L})$, and $p_{r\to m}= \hat{p}_{r\to m}/(1-\hat{p}_{r\to L})(1-\hat{p}_{r\to g})$, applied in that order.

\section{Detailed description of the error channel approximation}
\label{app:error_channel_approximation}

\renewcommand{\thetable}{B\arabic{table}}
\setcounter{table}{0}

\quad Each error channel defined in Appendix~\ref{app:noise_channel_definitions} requires a four- or six-dimensional Hilbert space to simulate the dynamics, which is computationally expensive. To trace the noise dynamics more efficiently, the noise channels are approximated to channels described by Eq.~\eqref{eq:error_channel_approximation} where the Pauli channel of Eq.~\eqref{eq:twirled_pauli_error_channel} and the erasure channel of Eq.~\eqref{eq:erasure_channel} act only within the computational subspace using the generalised Pauli twirling approximation (GPTA) of Section~\ref{subsec:approximation_to_pauli_noise_model}.
Then the noise events induced by these channels and noise propagation under the approximation can be traced through Clifford circuits, which is a natural way for implementing syndrome extraction circuits for stabiliser codes.

In this section, we describe the Clifford-simulatable form of each noise channel. For each channel we specify the erasure probability~$p$ of $\mathcal{E}_{\text{erase}}(\rho,p)$ and/or the probabilities $p_X$, $p_Y$, $p_Z$ that $\tilde{\mathcal{E}}_{\text{twirl}}$ assigns to $P_{\mu}=X,Y,Z$, respectively, composed as in Eq.~\eqref{eq:error_channel_approximation}. 
Section~\ref{app:approx_leakage_loss} describes how leakage and loss are realised in the stabiliser simulation, Sec.~\ref{app:approx_exact} lists the channels that require no approximation, and Secs.~\ref{app:approx_decay}--\ref{app:approx_measurement} present the approximated forms of the decay, clock-transition, and measurement channels.

\subsection{Leakage and loss as erasure}
\label{app:approx_leakage_loss}

\quad All loss channels and all decay channels whose error terms constitute leakage, i.e., transfer population out of the computational subspace, are replaced by the erasure channel $\mathcal{E}_{\text{erase}}(\rho,p)$ of Eq.~\eqref{eq:erasure_channel} with the corresponding transition probability~$p$. In the simulation, each such event is represented as a heralded-erasure instruction of \texttt{stim} with probability $p$. The qubit is replaced by the maximally mixed state, which is equivalent to applying the erasure channel in Eq.~\eqref{eq:erasure_channel}, and a herald flag is recorded but not supplied to the decoder.

This mapping relies on the following modelling assumptions. First, population that has left the computational subspace does not coherently return to the computational subspace, restricting the transition patterns to $c\rightarrow c$, $c\rightarrow e$, and $e\rightarrow e$ after all approximated noise channels. Second, the measurement of a lost atom returns a uniformly random outcome rather than a deterministic dark state detection. The classical discrimination error for the qubit states is contained in the $\mathrm{MERR}$ channel. Finally, the herald flags attached to the erasure channels on \texttt{stim} are assumed not to be available in our setting since obtaining this information physically would require additional leakage-detection or erasure-conversion operations~\cite{Wu2022-uy,Sahay2023-zo} that are not included in our device model. Accordingly, the heralded flags are not supplied to the decoder. Exploiting heralded erasures through erasure-aware decoding is expected to improve LERs~\cite{Perrin2025quantumerror,Baranes2026}, so incorporating various detection schemes has been left as future work, as discussed in Sec.~\ref{sec:conclusion}.

\subsection{Unapproximated channels}
\label{app:approx_exact}

\quad Pauli channels are invariant under Pauli twirling and can be inserted directly into the Clifford circuit simulation. The depolarising channels $\mathrm{DEP1}_{c}$, $\mathrm{DEP1}_{g}$, $\mathrm{DEP1}_{m}$, and $\mathrm{DEP2}_{c/m/\mathrm{dual}}$, the bit-flip channels $\mathrm{FLIP}_{g}$, $\mathrm{FLIP}_{m}$, $\mathrm{XERR}_{g}$, and $\mathrm{XERR}_{m}$, and the phase-flip channels $\mathrm{ZERR}_{c}$, $\mathrm{ZERR}_{gm}$, $\mathrm{ZERR}_{g}$, and $\mathrm{ZERR}_{m}$ act entirely within the computational subspace of the respective encoding. They are therefore represented without approximation, with the probabilities listed in Table~\ref{tab:noise_channel_summary}, evaluated at the operation times of Table~\ref{tab:gate_idling_times} for the time-dependent channels.

\subsection{Decay channels}
\label{app:approx_decay}

\quad The representation of the decay channels depends on whether the decay connects two computational states or leaves the computational subspace, and thus depends on the qubit encoding.

\medskip\noindent{\bfseries\boldmath $\mathrm{DECAY}_{mg}$ ($\boson$)} for the clock qubit is a transition $\ket{1}\rightarrow\ket{0}$ within the computational subspace, i.e., amplitude damping. The twirled amplitude damping channel with decay rate $\Gamma_{mg}$ and dephasing error $\mathrm{ZERR}_{c}$ with dephasing time $T_2$ are given by the probabilities described in Ref.~\cite{Tomita2014},
\begin{equation}
\begin{aligned}
  p_X = p_Y &= \frac{p_{m\rightarrow g}(t)}{4},\\
  p_Z &= \frac{p^{(c)}_{Z}(t)}{2} -\frac{p_{m\rightarrow g}(t)}{4},
\end{aligned}
\label{eq:app:ad_twirl}
\end{equation}
where $p_{m\rightarrow g}(t)=1-e^{-\Gamma_{mL}t}$ and $p^{(c)}_{Z}(t)=1-e^{-t/T_2}$.

\par\medskip\noindent{\bfseries\boldmath $\mathrm{DECAY}_{mg}$, $\mathrm{DECAY}_{mg}^{(\mathrm{gate})}$ ($\fermi$).} Both error channels for the $\fermi$-m qubit constitute decay out of the computational subspace. Both channels are therefore approximated as erasure channels, $\mathcal{E}_{\text{erase}}\bigl(\rho,p_{m\rightarrow g}(t)\bigr)$ and $\mathcal{E}_{\text{erase}}\bigl(\rho,p_{m\rightarrow g}^{(\mathrm{gate})}\bigr)$, respectively.
For the $\fermi$-g qubit, on the other hand, we assume these channels act as the identity because the decay from the metastable manifolds to the ground manifolds is negligible, except when applying the clock-transition excitation during the two-qubit gate.

\par\medskip\noindent{\bfseries\boldmath $\mathrm{DECAY}_{rg}$, $\mathrm{DECAY}_{rm}$, $\mathrm{LOSS}_{r}$.} The Rydberg decay channels act only during two-qubit gates. We assume that the population of the computational state coupled to the Rydberg state ($\ket{1}$ for $\boson$ and $\ket{m_1}$ for $\fermi$\footnote{Note that we assume $\fermi$-g qubits are driven to the metastable manifold by the clock-transition excitation in Appendix~\ref{app:approx_clock} before and after the CZ gate.}) is completely excited to $\ket{r}$ during the gate duration $t_{2Q}$, and the resulting single-qubit noise channels are applied to both atoms acted upon by the CZ gate.
Under this assumption, $\mathrm{LOSS}_{r}$ becomes an erasure process represented by $\mathcal{E}_{\text{erase}}\bigl(\rho,p_{r\rightarrow L}(t_{2Q})\bigr)$. Meanwhile, $\mathrm{DECAY}_{rm}$ returns the population to the computational subspace, destroying the phase coherence, which can be treated as a phase-flip error, $(p_X,p_Y,p_Z)=\bigl(0,0,p_{r\rightarrow m}(t_{2Q})\bigr)$. The effect of $\mathrm{DECAY}_{rg}$ depends on the encoding. For the clock qubit, it maps $\ket{1}$ onto $\ket{0}$ and is modelled as a bit-flip error, $(p_X,p_Y,p_Z)=\bigl(p_{r\rightarrow g}(t_{2Q}),0,0\bigr)$. In contrast, for the $\fermi$-m qubit, the ground manifold is outside the computational subspace. Thus $\mathrm{DECAY}_{rg}$ constitutes decay outside of the computational space and is assumed to be an erasure channel too.
Note that, since the $\fermi$-g qubit is temporarily transferred to the metastable manifold during two-qubit gates by the clock-transition excitation, $\fermi$-g qubits are also exposed to all the same Rydberg decay channels as $\fermi$-m qubits.

\subsection{Clock-transition excitation}
\label{app:approx_clock}

\quad While $\mathrm{DEP1}_{c}$ acts as the depolarising error channel for $\boson$ qubits, two-qubit gates and shelving operations on the $\fermi$-g qubit require a round trip $\ket{g_j}\rightarrow\ket{m_j}\rightarrow\ket{g_j}$ via the clock-transition. For each excitation or de-excitation, the depolarising error $\mathrm{DEP1}_{gm}$ on the $\fermi$-g qubit is approximated as erasure and Pauli channels from the error probability $p_{1}^{(gm)}$.
The GPTA of the channel $\mathrm{DEP1}_{gm}$ for the $\fermi$-g qubit yields an approximated error channel with $(p_X,p_Y,p_Z) = (0,0,p_{1}^{(gm)})$ and erasure probability $p_{\mathrm{erase}} = (2/3)\;p_{1}^{(gm)}$.
In addition, the metastable decay $\mathrm{DECAY}_{mg}$ acting during the clock-transition $t_{1Q}^{(gm)}$ is modelled as the composition of a bit-flip channel and an erasure channel with $(p_X,p_Y,p_Z)=\bigl(p_{m\rightarrow g}(t_{1Q}^{(gm)}),0,0\bigr)$ and $p_{\mathrm{erase}} = p_{m\rightarrow g}(t_{1Q}^{(gm)})$. 

For $\fermi$-m qubits, $\mathrm{DEP1}_{gm}$ accompanies the clock de-excitation step of the readout sequence described in Eq.~\eqref{eq:app:measurement_order_171m}. Although it contributes to readout error, we approximate the overall readout error with GPTA \emph{after} composing all of the error channels in~\eqref{eq:app:measurement_order_171m} together. Hence, we describe and account for it in Appendix~\ref{app:approx_measurement} rather than here.

\subsection{Measurement channels}
\label{app:approx_measurement}

\quad The composition of readout error is different for each qubit encoding as described in Sec.~\ref{app:measurement_channels}. Each noise channel is composed with the discrimination error channel $\mathrm{MERR}$ and quantum noise channels. In this section, we describe the explicit GPTA of each measurement error channel in Sec.~\ref{app:measurement_channels}.

\par\medskip\noindent{\bfseries\boldmath $\mathrm{MERR}$ ($\boson$)}\enspace---\enspace The GPTA of the measurement error $\mathrm{MERR}$ for $\boson$ qubits defined in Appendix~\ref{app:measurement_channels} yields the erasure probability of $\mathcal{E}_{\text{erase}}$ to be $p_{\mathrm{meas}}(1-p_{\mathrm{meas}})$, and the probabilities of each Pauli error on $\tilde{\mathcal{E}}_{\text{twirl}}$ with Pauli probabilities
\begin{equation}
\begin{aligned}
  p_X = p_Y &= \frac{1}{4}\,p_{\mathrm{meas}}+\frac{1}{2}\,p_{\mathrm{meas}}^{2} + \mathcal{O}(p_{\mathrm{meas}}^{3}),\\
  \textrm{and} \;p_Z &= \frac{q_{\mathrm{BB}}^{2}}{16}\,p_{\mathrm{meas}}^{2} + \mathcal{O}(p_{\mathrm{meas}}^{3}),
\end{aligned}
\label{eq:app:merr_twirl}
\end{equation}
where $q_{\mathrm{BB}}$ is the ratio for assigning the ambiguous bright-bright event to the $\ket{0}$ state or otherwise to the $\ket{1}$ state. We use $q_{\mathrm{BB}}=1/2$ as a default value. Terms of third order or higher in $p_{\mathrm{meas}}$ are neglected in our noise model since $p_{\mathrm{meas}} \ll 1$, so their contribution is negligible.

\par\medskip\noindent{\bfseries\boldmath $\mathrm{MERR}$, $\mathrm{FLIP}_{g}$, $\mathrm{DECAY}_{mg}$ during readout ($\fermi$-g)}\enspace---\enspace For the $\fermi$-g qubit, the bit-flip channel $\mathrm{FLIP}_{g}$ is a Pauli channel and requires no approximation. The discrimination error is represented by $\mathcal{E}_{\text{erase}}\bigl(\rho,\tfrac{3}{4}p_{\mathrm{meas}}\bigr)\circ\tilde{\mathcal{E}}_{\text{twirl}}$ with $(p_X,p_Y,p_Z)=\bigl(\tfrac{1}{4}p_{\mathrm{meas}},\tfrac{1}{4}p_{\mathrm{meas}},p_Z\bigr)$, where $p_Z=O(p_{\mathrm{meas}}^{2})$ is negligible at the rates considered here. We merge the erasure term from discrimination error with the readout loss $\mathrm{LOSS}_{g}^{(\mathrm{meas})}$, creating a single erasure channel. Furthermore, the state-selective imaging temporarily shelves one nuclear-spin population into the metastable manifold. Any population that decays back to the ground manifold during the readout window is assumed to land in either nuclear-spin state with equal probability, which corresponds to a bit-flip error at half the decay probability, $(p_X,p_Y,p_Z)=\bigl(p_{m\rightarrow g}(t_{\mathrm{read}})/2,0,0\bigr)$.

\par\medskip\noindent{\bfseries\boldmath Composite readout channel ($\fermi$-m)}\enspace---\enspace The readout sequence of $\fermi$-m qubits combines clock de-excitation with $\mathrm{DEP1}_{gm}$, dephasing and metastable-state decay during the readout time, state-selective imaging with $\mathrm{FLIP}_{g}$ and with $\mathrm{LOSS}_{g}^{(\mathrm{meas})}$, and the classical discrimination error (cf.~Eq.~\eqref{eq:app:measurement_order_171m}). Rather than twirling each factor separately, we apply the GPTA to the entire sequence and represent it by a single composition $\mathcal{E}_{\text{erase}}(\rho,p_{\mathrm{erase}})\circ\tilde{\mathcal{E}}_{\text{twirl}}$, as in Eq.~\eqref{eq:error_channel_approximation}, which is, to first order in the input probabilities,
\begin{equation}
\begin{aligned}
  p_{\mathrm{erase}} &= p_{\mathrm{meas}}+\tfrac{2}{3}p_{1}^{(gm)}+p_{g\rightarrow L}^{(\mathrm{meas})}+p_{m\rightarrow g}(t_{\mathrm{read}}),\\
  p_X &= \tfrac{1}{4}\,p_{\mathrm{meas}}+p_{\mathrm{flip}}^{(g)},\\
  p_Y &= \tfrac{1}{4}\,p_{\mathrm{meas}},\\
  p_Z &= p_{1}^{(gm)}+p_{Z}^{(m)}(t_{\mathrm{read}}),
\end{aligned}
\label{eq:app:171m_readout_twirl}
\end{equation}
where $p_{Z}^{(m)}(t)=1-e^{-t/T_{2}^{(m)}}$ and $p_{m\rightarrow g}(t)=1-e^{-\Gamma_{mg}t}$. Our simulation uses a closed-form expression that also includes the second-order terms of the six input probabilities, which has been validated against the direct numerical GPTA of the full sequence to within $10^{-6}$ over the parameter range considered in this work.

\section{Noise-Parameter Sensitivity of $\Lambda^{-1}$ in the single-isotope configurations}
\label{app:effects_of_error_reduction_for_Lambda}

\setcounter{figure}{0}
\renewcommand{\thefigure}{C\arabic{figure}}

\begin{figure*}
  \centering
  \includegraphics[width=1.0\textwidth]{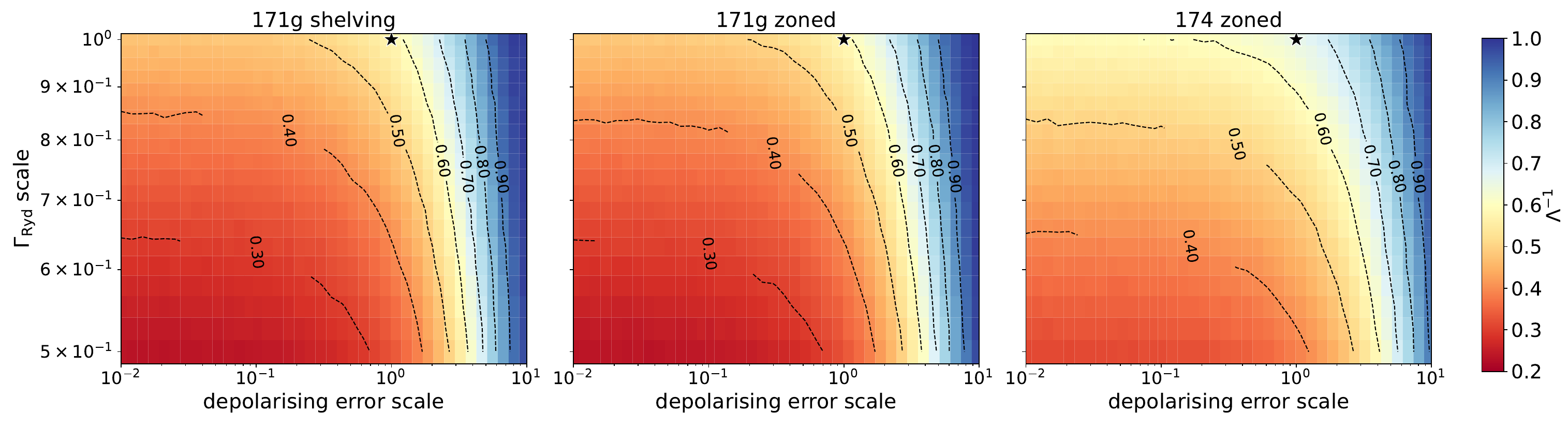}
  \caption{Error suppression rates for single-isotope configurations (ground-state and clock qubit) on the rotated surface code. Each heatmap shows the $\fermi$-g qubit with shelving, $\fermi$-g qubit with zoned measurement, and $\boson$ clock qubit with zoned measurement. The assumed noise parameters are the same as for the dual-isotope result in Fig.~\ref{fig:heatmap_lambda_inv_dual}. This figure indicates that the trends in $\Lambda^{-1}$ reduction are not changed significantly by the configuration of isotopes or the measurement method. Nevertheless, there is a small difference in the achievable value of $\Lambda^{-1}$ in each configuration. While configurations using $\fermi$-g qubits achieve almost the same performance as each other, the $\boson$ clock qubit shows a slightly worse $\Lambda^{-1}$.}
  \label{fig:app:heatmap_lambda_inv_single_isotope}
\end{figure*}

In this section, we show how $\Lambda^{-1}$ varies over different noise parameter regimes for single-isotope configurations, similar to the considerations found in the main text for the dual-isotope configuration presented in Fig.~\ref{fig:heatmap_lambda_inv_dual}.
We perform similar calculations here for the single-isotope cases; Fig.~\ref{fig:app:heatmap_lambda_inv_single_isotope} shows the heatmaps for these other configurations. The code and assumed noise parameters are the same as used in the dual-isotope results of Fig.~\ref{fig:heatmap_lambda_inv_dual}.
Pertinently, the trends of $\Lambda^{-1}$ over the decay rate $\Gamma_{\text{Ryd}}$ and the depolarising error rate remain the same, although there are small differences in the explicit values.
$\Lambda^{-1}$ in all configurations is more sensitive to a reduction in $\Gamma_{\text{Ryd}}$ than the depolarising error rate. Thus, we generally expect improvements in the Rydberg decay rate, or reducing the two-qubit gate time, to be effective at improving the QEC performance in single-isotope cases also.

Although there is similarity in the trends for decreasing $\Lambda^{-1}$, the achievable values for $\Lambda^{-1}$ are different in each configuration. While the three configurations in Figs.~\ref{fig:app:heatmap_lambda_inv_single_isotope} and ~\ref{fig:heatmap_lambda_inv_dual} ($\fermi$-g qubit with shelving, $\fermi$-g qubit with zoned measurement, and the dual-isotope configuration) show similar trends, the single-isotope configurations using $\boson$ clock qubits achieve a slightly worse $\Lambda^{-1}$ over the entire parameter regime.
Thus, using the $\fermi$-g qubit as the data qubit is expected to achieve better QEC performance.

\bibliographystyle{quantum}
\bibliography{Refs}

\end{document}